\PassOptionsToPackage{dvipsnames,svgnames,x11names}{xcolor}
\documentclass[preprintnumbers,amsmath,amssymb,10pt,prd,onecolumn,superscriptaddress,nofootinbib]{revtex4}
\usepackage{bbm}
\usepackage{amsfonts}
\usepackage{mathrsfs}
\usepackage{tikz}
\usepackage{latexsym}
\usepackage{epsfig}
\usepackage{epstopdf}
\usepackage{graphicx}
\usepackage{amssymb}
\usepackage{amsmath}
\usepackage{dcolumn}
\usepackage{bm}
\usepackage{comment}
\usepackage{multirow}
\usepackage{hhline}
\usepackage{array}
\usepackage{float}

\usepackage{booktabs}
\usepackage{makecell}
\newcolumntype{P}[1]{>{\centering\arraybackslash}p{#1}}

\begin{document}
\newcommand{\greenline}{\raisebox{2pt}{\tikz{\draw[-,black!40!green,solid,line width = 0.9pt](0,0) -- (5mm,0);}}}
\newcommand{\yellowline}{\raisebox{2pt}{\tikz{\draw[-,black!40!yellow,solid,line width = 0.9pt](0,0) -- (5mm,0);}}}
\newcommand{\orangeline}{\raisebox{2pt}{\tikz{\draw[-,black!40!orange,solid,line width = 0.9pt](0,0) -- (5mm,0);}}}
\newcommand{\purpleline}{\raisebox{2pt}{\tikz{\draw[-,black!40!purple,solid,line width = 0.9pt](0,0) -- (5mm,0);}}}
\newcommand{\Blueline}{\raisebox{2pt}{\tikz{\draw[-,black!40!NavyBlue,solid,line width = 0.9pt](0,0) -- (5mm,0);}}}
\newcommand{\cyanline}{\raisebox{2pt}{\tikz{\draw[-,black!40!cyan,solid,line width = 0.9pt](0,0) -- (5mm,0);}}}
\newcommand{\grayline}{\raisebox{2pt}{\tikz{\draw[-,black!40!gray,solid,line width = 0.9pt](0,0) -- (5mm,0);}}}
\newcommand{\blueline}{\raisebox{2pt}{\tikz{\draw[-,blue,dashed,line width = 1.5pt](0,0) -- (10mm,0);}}}
\newcommand{\magentaline}{\raisebox{2pt}{\tikz{\draw[-,magenta,dashed,line width = 1.5pt](0,0) -- (10mm,0);}}}

\title{\bf Charged Anisotropic Compact Stars in $f(R,\phi,X)$ Gravity: A Class I Embedding Approach with Reissner-Nordstr\"{o}m Exterior}

\author{Adnan Malik}
\email{adnan.malik@skt.umt.edu.pk}
\affiliation{School of Nuclear Science and Technology, University of South China, Hengyang 421001, China}
\affiliation{School of Mathematics and Physics, University of South China, Hengyang, 421001, China.}
\affiliation{Department of Mathematics, University of Management and Technology, Sialkot Campus, Pakistan.}
\affiliation{Center for Theoretical Physics, Khazar University, 41 Mehseti Str., AZ1096 Baku, Azerbaijan}

\author{Fatemah Mofarreh}
\email{Corresponding Author: fyalmofarrah@pnu.edu.sa}
\affiliation{Mathematical Science Department, Faculty of Science, Princess Nourah Bint Abdulrahman University, Riyadh 11546, Saudi Arabia}

\author{Ayesha Almas}
\email{ayeshaalmas852@gmail.com}
\affiliation{Department of Mathematics, University of Management and Technology, Sialkot Campus, Pakistan.}

\author{Wedad Albalawi}
\email{wsalbalawi@pnu.edu.sa}
\affiliation{Mathematical Science Department, Faculty of Science, Princess Nourah Bint Abdulrahman University, Riyadh 11546, Saudi Arabia}

\author{Aishah Alshehri}
\email{Aaalshehre@pnu.edu.sa}
\affiliation{Mathematical Science Department, Faculty of Science, Princess Nourah Bint Abdulrahman University, Riyadh 11546, Saudi Arabia}

\begin{abstract}
\begin{center}
\textbf{Abstract}
\end{center}

This work examines the physical characteristics of charged, anisotropic compact spheres within the framework of $f(R,\phi,X)$ modified gravity. Starting from a static, spherically symmetric spacetime, we employ an Adler-type ansatz for the temporal metric component $g_{tt}$. The corresponding radial metric component $g_{rr}$ is then systematically derived through application of the Karmarkar condition, which ensures a class-one embedding for the interior geometry. A crucial aspect of our approach involves matching the interior solution at the stellar boundary to the exterior Reissner--Nordstr$\ddot{0}$m spacetime---the established vacuum solution for charged, non-rotating masses. This matching procedure is essential for determining integration constants and verifying global physical consistency. Our comprehensive analysis of the resulting stellar model investigates multiple physical aspects, including energy density, radial and tangential pressures, anisotropy, equation-of-state parameters, energy conditions, mass function, compactness, and surface redshift. Stability assessment further incorporates examination of the adiabatic index and the Tolman--Oppenheimer--Volkoff equation. Collectively, our findings demonstrate that the charged compact star model presented here constitutes a physically viable configuration---free of singularities and maintaining stable equilibrium across the considered parameter space.
\end{abstract}

\maketitle

\date{\today}

\section{Introduction}

Compact stellar objects---including neutron stars, quark stars, and black holes---continue to captivate researchers by offering unique laboratories for probing fundamental physics under extreme gravitational conditions. While Einstein's general relativity has long served as the cornerstone for understanding these systems, its inability to fully account for cosmic acceleration and dark energy has spurred interest in alternative gravitational theories \cite{1b,3b}. The $f(R)$ approach, initiated by Buchdahl \cite{4b}, extends the Einstein--Hilbert action in a natural way and has played a key role in building realistic dark energy models \cite{5b}. Further developments, including Harko and collaborators' $f(R,T)$ theory \cite{7b}, incorporate direct curvature--matter interactions, giving rise to non-conservation of the stress--energy tensor and additional non-geodesic accelerations. Some other modified theories of gravities such as $f(R)$, $f(G)$, $f(R,G)$, $f(T)$, $f(R, T)$, $f(Q)$ etc, which introduce richer geometric structure through higher-order curvature terms or non-minimal matter couplings, thereby expanding the landscape of possible compact object interiors \cite{11b}. These advances have significantly broadened the scope for studying anisotropic stellar configurations and their stability, deepening our insight into gravitational collapse and compact star formation \cite{12b}. Introducing electric charge into compact stellar models profoundly influences their equilibrium and stability, reshaping the collapse dynamics in essential ways. A charged fluid source produces a repulsive electromagnetic force that can oppose gravitational pull, potentially arresting collapse and avoiding spacetime singularities \cite{14b,15b}. This prospect has motivated substantial work on charged compact stars in modified gravity, where strengthened gravitational coupling can combine with electromagnetic repulsion to yield stable, regular configurations. For example, Malik et al. \cite{15ba} showed that within $f(R,\phi,X)$ gravity, charged anisotropic spheres display a rich set of physical properties, meeting all standard energy conditions and stability bounds.

Discussion of stellar structures through Karmarkar condition is very attractive topic among the researchers. Karmarkar \cite{19b} demonstrated that while general spherically symmetric gravitational metrics are typically of class two, specific conditions can reduce them to class one. Prasad et al. \cite{a1} obtained some families of relativistic anisotropic compact stars by solving of Einstein's field equations and Karmarkar condition. Sharif and Gul \cite{a3} examined the viability and stability of anisotropic compact stellar objects adopting the Karmarkar condition in energy-momentum squared gravity. Bhar et al. \cite{a5} presented a comparative study on generalized model of anisotropic compact star satisfying the Karmarkar condition. Singh et al. \cite{a6} presented a new class of solutions to the Einstein field equations for an anisotropic matter distribution in which the interior space-time obeys the Karmarkar condition. Abbas et al. \cite{a7} investigated the anisotropic compact stars using quintessence and Karmarkar conditions.

Solving the field equations in modified gravity often presents substantial mathematical challenges, prompting the use of specific metric ansatzes that preserve physical relevance while simplifying the analysis. In this context, the Adler--Finch--Skea metric has proven especially useful, supplying a tractable setting for studying static, spherically symmetric solutions \cite{21b,22b}. This metric yields solutions that are free of singularities at the origin and produce well-behaved density and pressure profiles, making it particularly suitable for modeling compact stars \cite{23b,24b}. When used alongside the Karmarkar condition for embedding class one spacetimes, the Adler--Finch--Skea ansatz supports the construction of physically realistic anisotropic fluid spheres that meet standard stability requirements \cite{25b,26b}. Recent applications in modified gravity highlight its effectiveness in generating solutions that capture the complex internal structure of compact stars. The combination of mathematical simplicity and physical soundness has made this approach a common tool in relativistic astrophysics, allowing researchers to explore compact objects beyond the restrictions of simple perfect fluid models \cite{30b}.

The Reissner--Nordstr$\ddot{0}$m solution, which describes a charged, non-rotating black hole in general relativity, also provides an important reference for analyzing compact stellar structures in extended gravity settings. As an exact electrovacuum solution, it describes the exterior geometry of a static, charged, spherical mass and serves as a key standard for matching interior solutions \cite{31b,32b}. When modeling charged compact stars in modified gravity, the Reissner--Nordstr$\ddot{0}$m metric often plays the role of the exterior spacetime, allowing one to fix boundary conditions that smoothly join the interior and exterior geometries \cite{33b,34b}. This matching process is especially relevant in $f(R,\phi,X)$ gravity, where extra scalar and kinetic terms alter the junction conditions at the stellar surface \cite{35b,36b}. Investigations using the Reissner--Nordstr$\ddot{0}$m exterior have shed light on how electric charge affects maximum mass limits, surface redshifts, and stability boundaries of compact objects in modified gravity settings \cite{37b,38b}. Moreover, this methodology has been valuable in studying regular black hole spacetimes, such as the Bardeen metric, which can be viewed as arising from nonlinear electrodynamics with magnetic monopole sources \cite{r31,r33,r34,r35}.

Modified theories of gravity especially $f(R,\phi,X)$ theory of gravity is a good candidate to address both dark energy and compact object structure. By incorporating a scalar field $\phi$ and its kinetic term $X$ together with the Ricci scalar $R$, this theory spans a broad class of dark energy models and extended gravity formulations. Bahamonde and his collaborators \cite{13ba} used reconstruction methods to determine specific $f(R,\phi,X)$ actions consistent with the observed cosmic acceleration, confirming its viability as a dark energy model. Recently, Li et al., \cite{41b} discussed the wormhole solutions in the presence of charge in modified $f(R,\phi,X)$ theory of gravity. Shamir along with his collaborators \cite{43b} investigated the traversable wormhole solutions and non-commutative wormhole solutions in the same modified theory of gravity. Sharif and Gul \cite{44b} investigated the behavior of anisotropic stellar structures admitting Karmakar condition in modified $f (R, \phi, X)$ modified theory of gravity.

It is important to contextualize the novelty of our approach within the existing literature. While the Adler--Finch--Skea ansatz, the Karmarkar embedding condition, and matching to the Reissner--Nordstr{\"o}m exterior are established tools in relativistic astrophysics \cite{21b, finch1989}, their application within the $f(R,\phi,X)$ framework yields new physical insights that are not simply a repetition of previous work. The unique structure of $f(R,\phi,X)$ gravity introduces a direct coupling between the scalar field kinetic term $X$ and the fluid anisotropy, which, in the presence of an electric field, leads to qualitatively different behavior than in GR or simpler $f(R)$ models. Furthermore, our model makes testable predictions; we find that the curvature corrections systematically enhance the stellar compactness and surface redshift compared to GR predictions for stars of the same mass and charge. Quantifying these deviations is a central aim of this work and provides a potential pathway for observationally distinguishing modified gravity from GR using compact star data.
In contrast, $f(R,\phi,X)$ gravity provides a unified framework that combines: (i) curvature corrections via $R$ and $R^{2}$ terms, (ii) a dynamical scalar field $\phi$ with its own potential $V(\phi)$, and (iii) a kinetic term $X = -\frac{1}{2}g^{\mu\nu}\nabla_{\mu}\phi\nabla_{\nu}\phi$ that couples directly to the metric. This structure yields two key advantages: the scalar field can model anisotropic pressure without invoking exotic matter, and the kinetic term introduces a non-minimal coupling that modifies the TOV equation in ways absent in simpler theories \cite{bahamonde2019, malik2022}. Importantly, our model encompasses several limiting cases: setting $\tau = 0$ and $\phi = \text{constant}$ recovers GR; constant $\phi$ with $\tau \neq 0$ yields $f(R)$ gravity; and $\tau = 0$ with dynamical $\phi$ reduces to scalar-tensor theory. This hierarchical relationship allows us to isolate and quantify the specific contributions of each component to stellar structure.

At this juncture, it is pertinent to clarify why $f(R,\phi,X)$ gravity merits specific attention compared to other modified gravity formulations. While theories such as $f(R)$ \cite{nashed2021}, $f(T)$ \cite{malik2024teleparallel}, $f(R,T)$ \cite{maurya2017}, $f(G)$ \cite{naz2024fg}, and $f(R,G)$ \cite{naz2023frg} have each provided valuable insights into compact object physics, they possess inherent limitations that $f(R,\phi,X)$ overcomes. First, $f(R)$ gravity, despite its success in explaining cosmic acceleration, introduces only a single scalar degree of freedom and cannot independently model anisotropic effects without additional assumptions about the matter sector. Second, $f(T)$ gravity, while phenomenologically rich, suffers from the drawback of lacking local Lorentz invariance in its general formulation. Third, $f(R,T)$ gravity couples curvature directly to the trace of the energy-momentum tensor, but this coupling is algebraic rather than dynamical, limiting its ability to generate evolving anisotropic structures. Fourth, $f(G)$ and $f(R,G)$ theories incorporate higher-order curvature invariants but often lead to complex field equations that obscure the physical interpretation of stellar interiors.

In this work, we develop a new model for charged anisotropic compact stars in $f(R,\phi,X)$ gravity, combining the Adler--Finch--Skea metric ansatz with the Reissner--Nordstr$\ddot{0}$m exterior geometry. Our model builds on the Karmarkar condition for embedding class one spacetimes to obtain exact interior solutions, which we then scrutinize using a full set of physical diagnostics: energy conditions, stability tests, and mass--radius relations. The structure of the paper is as follows. Section~2 lays out the theoretical foundations of $f(R,\phi,X)$ gravity and presents the field equations for a charged, anisotropic fluid. Section~3 applies the Adler--Finch--Skea metric within the embedding class one framework and details the matching to the external Reissner--Nordstr$\ddot{0}$m geometry. Section~4 carries out a comprehensive physical analysis of the resulting solutions, examining density, pressure, anisotropy, energy conditions, and stability through multiple criteria. Finally, Section~5 summarizes our main results and considers their implications for modeling compact stars in modified gravity.

\section{Basic Formulism of $f(R,\phi,X)$ Modified Gravity }

The action for the $f(R,\phi,X)$ theory of gravity is defined \cite{c1,c2} as
\begin{equation}\label{1}
S = \int d^4x \sqrt{-g} \left( \frac{1}{2\kappa} f(R,\phi,X) + \mathcal{L}_m \right),
\end{equation}
where $g$ is the determinant of the metric tensor $g_{\eta\xi}$, $\mathcal{L}_m$ is the matter Lagrangian, and $f(R,\phi,X)$ is an arbitrary function of the Ricci scalar $R$, a scalar field $\phi$, and its kinetic term $X = -\frac{1}{2} (\nabla \phi)^2$. Varying the action with respect to the metric tensor yields the following field equations as
\begin{equation}\label{3}
\begin{split}
f_R G_{\xi\eta} - \frac{1}{2}(f - R f_R) g_{\xi\eta} - \nabla_\xi \nabla_\eta f_R + g_{\xi\eta} \nabla_\nu \nabla^\nu f_R - \frac{1}{2} f_X (\nabla_\xi \phi)(\nabla_\eta \phi) = \kappa T_{\xi\eta},
\end{split}
\end{equation}
where $f_R \equiv \partial f / \partial R$, $f_X \equiv \partial f / \partial X$, and $\nabla_\nu$ denote the covariant derivative. Moreover, we consider the energy-momentum tensor in the presence of charge as
\begin{equation}\label{4}
T_{\xi\eta} = (\rho + p_t) \zeta_{\xi} \zeta_{\eta} - p_t g_{\xi\eta} + (p_r - p_t) \vartheta_{\xi} \vartheta_{\eta} + \frac{1}{4\pi} \left( F_{\xi}^{\mu} F_{\eta\mu} - \frac{1}{4} g_{\xi\eta} F_{\mu\nu} F^{\mu\nu} \right),
\end{equation}
where $\rho$ is the energy density, $p_r$ and $p_t$ are the radial and tangential pressures, respectively. The four-velocity $\zeta_{\eta}$ and radial unit vector $\vartheta_{\xi}$ are defined as $\zeta_{\eta} = e^{\lambda/2} \delta^0_\eta$ and $\vartheta_{\xi} = e^{\beta/2} \delta^1_\xi$. The electromagnetic field tensor $F_{\xi\mu}$ satisfies Maxwell's equations:
\begin{equation}\label{5}
F_{\xi\mu} = B_{\mu,\xi} - B_{\xi,\mu}, \quad {F^{\xi\mu}}_{;\mu} = -4\pi J^{\xi},
\end{equation}
where $B_{\mu}$ is the four-potential and $J^{\xi} = \sigma v^{\xi}$ is the four-current density, with $\sigma$ representing the charge density. We assume a static, spherically symmetric spacetime described by the line element:
\begin{equation}\label{6}
ds^{2} = e^{\lambda(r)} dt^{2} - e^{\beta(r)} dr^{2} - r^{2} (d\theta^{2} + \sin^2\theta \, d\phi^{2}).
\end{equation}
For this metric, the only non-vanishing component of the Maxwell tensor is $F^{01} = -F^{10}$, given by:
\begin{equation}\label{7}
F^{01} = \frac{q}{r^2} e^{-(\lambda + \beta)/2},
\end{equation}
where $q(r)$ represents the total charge within a radius $r$:
\begin{equation}\label{8}
q(r) = 4\pi \int_0^r \sigma(\rho) \rho^2 e^{\beta(\rho)/2} d\rho.
\end{equation}
The electric field intensity $E$ is consequently:
\begin{equation}\label{9}
E^2 = -F^{10}F_{01} = \frac{q^2}{r^4}.
\end{equation}
Substituting the metric (\ref{6}) and the energy-momentum tensor (\ref{4}) into the field equations (\ref{3}), we get
\begin{equation}\label{10}
\begin{split}
\rho + E^2 =& \frac{1}{4r e^\lambda} \left( 2r\beta'' - \beta' - r\lambda'^2 + 4\beta' \right) f_R - \frac{f}{2} - e^{-\lambda} \left( \frac{\lambda'}{2} + \frac{2}{r} \right) f_R' - e^{-\lambda} f_R'',
\end{split}
\end{equation}
\begin{equation}\label{11}
\begin{split}
p_r - E^2 =& \frac{1}{4 e^\lambda} \left( -2r\beta'' - \beta' + \lambda'^2 + \frac{4\lambda'}{r} \right) f_R + \frac{f}{2} + e^{-\lambda} \left( \frac{\beta'}{2} + \frac{2}{r} + \lambda' \right) f_R' - \frac{1}{2e^{\lambda}} f_X \phi'^2,
\end{split}
\end{equation}
\begin{equation}\label{12}
\begin{split}
p_t + E^2 =& \frac{1}{4r^2 e^\lambda} \left( -r\beta' + r\lambda' + 2e^\lambda - 2 \right) f_R + \frac{f}{2} + e^{-\lambda} \left( \frac{\beta'}{2} + \frac{\lambda'}{2} + \frac{1}{r} + \lambda' \right) f_R' + e^{-\lambda} f_R'',
\end{split}
\end{equation}
\begin{equation}\label{13}
\sigma = \frac{e^{-\beta/2}}{4\pi r^2} (q r^2)'.
\end{equation}
Here, a prime $(^\prime)$ denotes differentiation with respect to the radial coordinate $r$. To close the system of equations, we impose the Karmarkar condition \cite{c3}, a necessary and sufficient condition for a spherically symmetric spacetime to be of embedding class one. This condition reads:
\begin{equation}\label{22a}
R_{2323} R_{1414} = R_{3434} R_{1212} + R_{1334} R_{1224}, \quad \text{with} \quad R_{2323} \neq 0.
\end{equation}
After substituting the values, this reduces to the following differential equation:
\begin{equation}\label{24a}
\frac{\beta' \lambda'}{1 - e^{\lambda}} - \left( \beta' \lambda' + \beta'^2 - 2(\beta'' + \beta'^2) \right) = 0, \quad \text{with} \quad e^\beta \neq 1.
\end{equation}
Solving Eq. (\ref{24a}), we get
\begin{equation}\label{25a}
e^{\beta(r)} = 1 + H e^{\lambda(r)} \lambda'(r)^2,
\end{equation}
where $H \neq 0$ is an arbitrary integration constant. We now adopt a specific ansatz for $g_{tt}$ metric potential, proposed by Adler \cite{c4} as
\begin{equation}\label{26a}
e^{\lambda(r)} = J (1 + L r^2)^2,
\end{equation}
where $J > 0$ and $L$ are constants. Substituting Eq. (\ref{26a}) into Eq. (\ref{25a}) yields the following expression for $g_{rr}$
\begin{equation}\label{27a}
e^{\beta(r)} = 1 + 16 J H L^2 r^2.
\end{equation}
This form is analogous to the well-known Finch--Skea solution \cite{c5}. The chosen metric potentials are well-behaved, non-singular at the origin, and monotonically increasing, satisfying the conditions for physical viability as discussed by Lake \cite{c6}. This particular combination of metric potentials has previously yielded successful compact star models in various modified gravity settings \cite{c7,c8}. At this juncture, it is important to provide the physical motivation for our specific choice of the temporal metric potential in Eq.~(\ref{26a}) and the resulting embedding class one geometry. The Adler ansatz $e^{\lambda(r)} = J(1+Lr^2)^2$ is not an arbitrary selection but is guided by several physical considerations. First, it ensures regularity at the center: $e^{\lambda(0)} = J$ (finite) and $e^{\lambda'(0)} = 0$, which guarantees that all curvature invariants remain finite at the core. Second, it reduces to the well-known Schwarzschild interior solution in the appropriate limit, thereby maintaining consistency with standard general relativistic results for uniform density stars \cite{schwarzschild1916}. Third, the potential is monotonically increasing with $r$, a necessary condition for a stable gravitational field configuration. When combined with the Karmarkar condition (\ref{24a}), this ansatz generates the radial metric potential $e^{\beta(r)} = 1 + 16JHL^2r^2$, which is of the Finch-Skea type \cite{finch1989}. The Finch-Skea metric has been extensively validated in the literature as producing well-behaved physical profiles: energy density and pressures are positive and monotonically decreasing, the sound speeds are subluminal, and the configuration satisfies all energy conditions \cite{23b}. Moreover, this metric form has been successfully applied to model compact stars in various modified gravity contexts, including $f(R,T)$ \cite{maurya2018} and $f(R,\phi)$ gravity \cite{malik2024}, demonstrating its versatility and physical robustness.

We also considered alternative metric potentials during the development of this work. Tolman-type solutions \cite{tolman1939}, where $e^{\lambda}$ is taken as constant or a simple power law, were examined but found to be too restrictive for charged anisotropic fluids, often leading to unphysical behavior in the pressure anisotropy or violating the causality condition. Buchdahl-type ansatze \cite{buchdahl1959}, while useful for certain isotropic configurations, introduce additional mathematical complexity that can obscure the physical interpretation of the scalar field and curvature corrections in $f(R,\phi,X)$ gravity. The Adler-Finch-Skea combination adopted here strikes an optimal balance between mathematical tractability and physical realism, enabling a clear and systematic analysis of the modified gravity effects on stellar structure.

For the electric field distribution, we adopt a functional form that maintains analytical tractability while capturing essential physical behavior \cite{c9}:
\begin{equation}\label{E_field_form}
E^2 = K L r,
\end{equation}
with $K$ representing a dimensionless constant. At this stage, a comment on the physical interpretation of the electric charge is warranted. Throughout this work, we adopt a representative charge-to-mass ratio of $Q/M = 0.3$ for all stellar candidates. While this value is relatively large compared to estimates of net charge in realistic neutron stars, it is important to clarify that this ``charge'' should not be interpreted literally as a net electrostatic charge on the stellar surface. Such a large net charge would be astrophysically untenable due to rapid discharge via the interstellar medium or pair production processes \cite{ray2003}. Instead, the charge parameter in our model serves two possible physical interpretations.

First, it can be understood as an \textit{effective charge} arising from the modified gravity sector. In $f(R,\phi,X)$ gravity, the non-minimal couplings between curvature, scalar field, and matter can produce contributions to the energy-momentum tensor that mimic electromagnetic effects \cite{bahamonde2019}. The geometric and scalar field terms in Eqs.~(\ref{10})--(\ref{12}) appear alongside the electromagnetic field tensor, making it difficult to disentangle literal charge from effective curvature-induced charge. Thus, our $Q/M = 0.3$ may be viewed as encoding the combined influence of electromagnetic fields and modified gravity. Second, such a value can represent the volume-averaged effect of ultra-strong magnetic fields. In magnetars, surface magnetic fields reach $10^{14}$--$10^{15}$ G, and theoretical models suggest that interior fields could be orders of magnitude higher \cite{duncan1992, thompson1994}. Through the equivalence $B^2 \sim Q^2/r^4$, our $Q/M = 0.3$ corresponds to an effective magnetic field strength of order $10^{17}$--$10^{18}$ G when averaged over the stellar volume. While extreme, such field strengths have been considered in studies of magnetized compact objects \cite{haensel2007, chatterjee2019} and remain within the bounds set by virial theorems. Therefore, our chosen $Q/M = 0.3$ is not intended as a precise astrophysical prediction but as a representative value within the theoretically allowed range that yields well-behaved solutions and clearly illustrates the effects of the electromagnetic coupling in $f(R,\phi,X)$ gravity. The qualitative features of our results are robust for a range of $Q/M$ values, and $0.3$ was selected to ensure these effects are discernible while remaining mathematically tractable.
For our current work, we consider $f(R,\phi,X)$ gravity model as
\begin{equation}\label{14c}
f(R,\phi,X) = R + \tau R^2 - V(\phi) + X.
\end{equation}
This model incorporates several physically meaningful components: a quadratic curvature correction $\tau R^2$ that can help resolve issues related to primordial singularities and early-universe inflation; a scalar potential $V(\phi)$; and the canonical kinetic term $X$. For the scalar potential, we adopt a power-law form:
\begin{equation}
V(\phi) = w_0 \phi^m,
\end{equation}
while the scalar field itself follows a radial power-law distribution:
\begin{equation}
\phi(r) = r^{\beta},
\end{equation}
where $w_0$, $m$, and $\beta$ are non-zero constants. The pressure anisotropy, defined through $\Delta = p_t - p_r$, follows directly from these expressions. Given the considerable complexity of Eqs. (\ref{15})-(\ref{17}), a graphical approach becomes essential for assessing the physical viability and stability characteristics of the resulting stellar configurations---an approach we pursue in the following sections.

\section{Boundary Conditions}
To determine the unknown parameters in our interior solution, we match it at the stellar boundary \( (r = R) \) with an appropriate exterior geometry. For a charged, non-rotating compact star, the correct exterior is given by the Reissner--Nordstr$\ddot{0}$m metric. Maintaining continuity between the interior and exterior geometries is essential for a physically meaningful model, as it ensures a smooth transition across the boundary surface. We take the Reissner--Nordstr$\ddot{0}$m metric as the exterior spacetime, expressed as:
\begin{equation}\label{18}
ds^{2} = u(r)dt^{2} - u(r)^{-1}dr^{2} - r^{2}(d\theta^{2} + \sin^2\theta \, d\phi^{2}),
\end{equation}
where the function \( u(r) \) is given by:
\begin{equation}\label{19}
u(r) = 1 - \frac{2M}{r} + \frac{Q^2}{r^2}.
\end{equation}
Here, \( M \) and \( Q \) represent the total gravitational mass and total charge of the stellar structure, as measured by an external observer. This metric is a vacuum solution to the Einstein--Maxwell equations and satisfies the singularity theorems, making it a consistent spacetime for the exterior of a charged sphere. At the stellar boundary \( (r = R) \), the interior metric must match the exterior Reissner--Nordstr$\ddot{0}$m metric smoothly. This requires the following junction conditions:
\begin{equation}\label{21}
 {g_{tt}}^{+}(R) = {g_{tt}}^{-}(R), \qquad {g_{rr}}^{+}(R) = {g_{rr}}^{-}(R), \qquad \left.\frac{\partial g_{tt}^{+}}{\partial r}\right|_{r=R} = \left.\frac{\partial g_{tt}^{-}}{\partial r}\right|_{r=R}.
\end{equation}
Here, the superscripts \( + \) and \( - \) denote the exterior and interior metrics, respectively. By using these junction conditions, we get the values of unknowns

\begin{equation}\label{26}
L = \frac{1}{2R^2} \left( \sqrt{\frac{R - 2M + \frac{Q^2}{R}}{R - 3M + \frac{2Q^2}{R}}} - 1 \right),
\end{equation}
\begin{equation}\label{27}
J = \frac{\left( R - 3M + \frac{2Q^2}{R} \right)^2}{R^2 \left( R - 2M + \frac{Q^2}{R} \right)},
\end{equation}
\begin{equation}\label{28}
H = \frac{R^3}{16 J L^2} \left( \frac{1}{R - 2M + \frac{Q^2}{R}} - 1 \right).
\end{equation}
The parameter \( \tau \) from the \( f(R,\phi,X) \) model is determined numerically by enforcing the boundary condition \( p_r(R) = 0 \) using the expressions for the interior solution. Next, we calculate the numerical values of these parameters. For this analysis, we assume a representative charge-to-mass ratio, \( Q/M = 0.3 \), and use the following fixed values for the other model parameters: \( m=0.1 \), \( w_0=1\times10^{-5} \), \( K=0.000167 \), and \( \beta=1\times10^{-6} \). The results are presented in Table~1.
\begin{table}[H]
\centering
\caption{Calculated parameters for the compact star candidates for \( Q/M = 0.3 \), \( m=0.1 \), \( w_0=1\times10^{-5} \), \( K=0.000167 \), and \( \beta=1\times10^{-6} \).}
\renewcommand{\arraystretch}{1.8}
\begin{tabular}{c c c c c c c}
\hline
\textbf{Compact Star Model} & $M/M_\odot$ & $R$ (km) & $L$ (km$^{-2}$) & $H$ & $J$ & $\tau$  \\
\hline
$\textbf{EXO 1785-248}~(CS_1)$ & 1.30   & 8.849 & 0.00521 & 154.32 & 0.452 & -0.0182 \\
$\textbf{Vela X - 1}~(CS_2)$ & 1.77  & 9.56 & 0.00688 & 142.15 & 0.285 & -0.0714 \\
$\textbf{Her X-1}~(CS_3)$ & 0.85  & 8.1 & 0.00315 & 198.74 & 0.624 & -0.00211 \\
$\textbf{LMC X-4}~(CS_4)$ & 1.29  & 9.711 & 0.00389 & 205.33 & 0.487 & -0.0128 \\
$\textbf{4U 1608}~(CS_5)$ & 1.74  & 9.3 & 0.00724 & 132.88 & 0.261 & -0.0721 \\
$\textbf{SAX J1808.4-3658}~(CS_6)$ & 0.9 & 7.951 & 0.00374 & 176.45 & 0.665 & -0.00398 \\
$\textbf{Cen X-3}~(CS_7)$ & 1.49  & 10.136 & 0.00412 & 201.56 & 0.421 & -0.0231 \\
$\textbf{PSR J1903+327}~(CS_8)$ & 1.667  & 9.82 & 0.00564 & 168.90 & 0.325 & -0.0451 \\
$\textbf{ PSR J1614-2230}~(CS_9)$ & 1.97  & 10.30 & 0.00705 & 159.24 & 0.234 & -0.1105 \\
\hline
\end{tabular}
\renewcommand{\arraystretch}{1}
\end{table}

\section{Physical Analysis and Visual Investigation}

In this section, we present a comprehensive physical analysis of our charged anisotropic compact star model within the $f(R,\phi,X)$ gravity framework. We begin by establishing the unit system used throughout. Subsequently, we systematically examine the behavior of key thermodynamic variables, the anisotropy factor, equation-of-state parameters, energy conditions, and stability criteria. For each physical quantity, we verify the necessary conditions for a realistic stellar configuration: regularity at the center, monotonic decrease toward the surface, positivity, and compliance with causality and energy bounds. The analysis is performed for nine well-known compact star candidates whose parameters are listed in Table~I, and all results are illustrated graphically to facilitate visual inspection and comparison across candidates.

\textbf{Unit System:} Throughout this work, we employ geometric units in which the gravitational constant and speed of light are set to unity, $G = c = 1$. In this system, all quantities with dimensions of length, mass, and time are expressed in kilometers (km). Specifically, energy density $\rho$ and pressures $p_r, p_t$ have units of km$^{-2}$. Conversion to physical cgs units (g/cm$^3$ and dyn/cm$^2$) can be performed using the relations $1\,\text{km}^{-2} = 1.801 \times 10^{17}\,\text{g/cm}^3$ and $1\,\text{km}^{-2} = 1.801 \times 10^{35}\,\text{dyn/cm}^2$, respectively. This convention ensures consistency with the metric potentials $e^{\lambda(r)}$ and $e^{\beta(r)}$, which are dimensionless.

This section presents a detailed examination of the physical characteristics of our anisotropic, charged compact star model, developed within the class-one embedded spacetime formalism. The parameters of the model depend explicitly on the stellar mass \( M \) and radius \( R \). To ensure astrophysical relevance, we have computed these parameters for a set of known compact star candidates, as summarized in Table~1. Using these results, we now assess the physical acceptability of the model via graphical methods.

A fundamental criterion for any physically realistic stellar model concerns the behavior of the key thermodynamic variables: the energy density \( \rho \), radial pressure \( p_r \), and tangential pressure \( p_t \). For the model to be admissible, these quantities must remain non-negative throughout the stellar interior. Each should attain its maximum at the center \( (r = 0) \) and decrease monotonically toward the surface. Furthermore, at the boundary \( (r = R) \), the radial pressure must vanish, \( p_r(R) = 0 \), while the energy density and tangential pressure should remain positive.

\begin{figure}[!ht]
    \centering
    \includegraphics[scale=0.8]{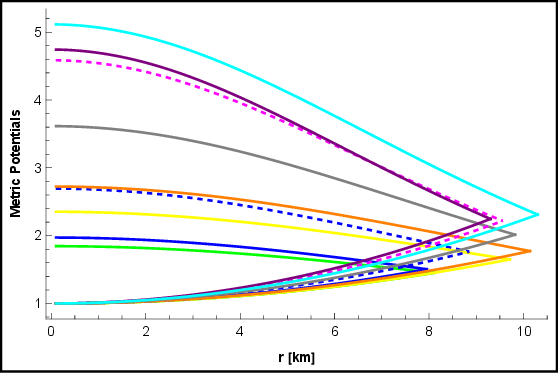}
    \includegraphics[scale=0.8]{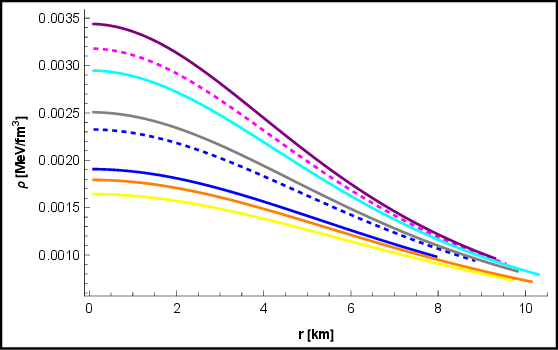}
    \caption{Graphical variation of $e^{\lambda}$, $e^{\beta}~and$ $\rho$ against $r$ for $CS_1$ (\protect\blueline), $CS_2$ (\protect\magentaline), $CS_3$ (\protect\greenline), $CS_4$ (\protect\yellowline), $CS_5$ (\protect\purpleline), $CS_6$ (\protect\Blueline), $CS_7$ (\protect\orangeline), $CS_8$ (\protect\grayline) and $CS_9$ (\protect\cyanline) against $r$.}
    \label{rho}
\end{figure}

\begin{figure}[!ht]
    \centering
    \includegraphics[scale=0.8]{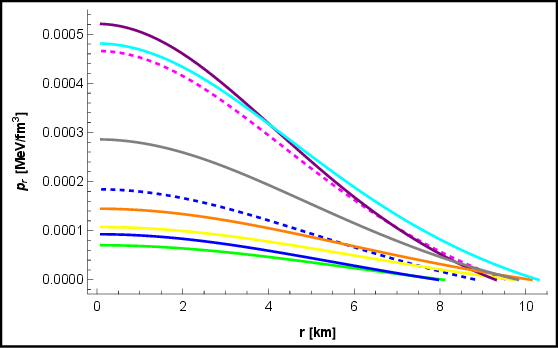}
    \includegraphics[scale=0.8]{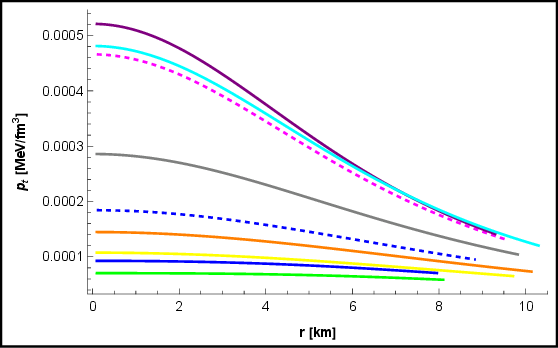}
    \caption{Graphical evaluation of $p_r~and~p_t$  for $CS_1$ (\protect\blueline), $CS_2$ (\protect\magentaline), $CS_3$ (\protect\greenline), $CS_4$ (\protect\yellowline), $CS_5$ (\protect\purpleline), $CS_6$ (\protect\Blueline), $CS_7$ (\protect\orangeline), $CS_8$ (\protect\grayline) and $CS_9$ (\protect\cyanline) against $r$.}
    \label{pr}
\end{figure}

The graphical behavior of these quantities, presented in Figs.~\ref{rho} and \ref{pr}, confirms that our model successfully satisfies all these essential criteria. The profiles demonstrate finite, positive values at the core and a smooth, decreasing trend outward, culminating in the required boundary condition for the radial pressure.

The central values of these quantities, which must be finite and positive, are provided by the expressions given in the Appendix. These central values are confirmed to be finite and positive within the considered parameter space, ensuring a regular and physically plausible core for the compact star.
\begin{figure}[!ht]
    \centering
    \includegraphics[scale=0.8]{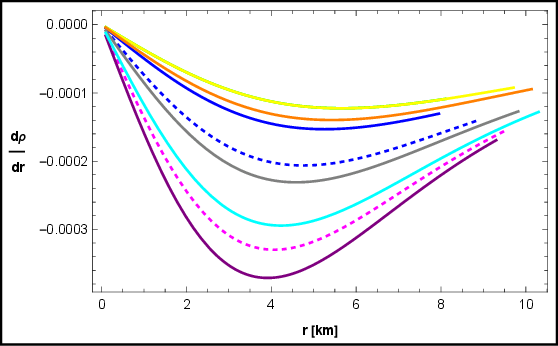}
    \includegraphics[scale=0.8]{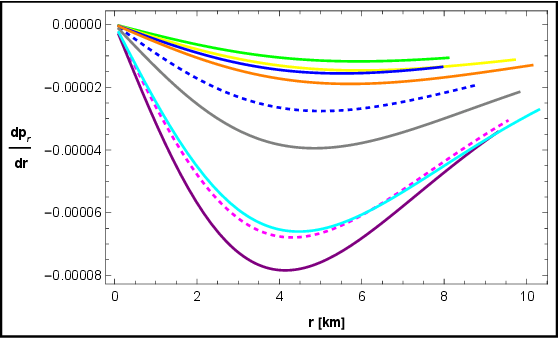}
    \caption{Variation of $\frac{dp_r}{dr}~and~\frac{d\rho}{dr}$ for $CS_1$ (\protect\blueline), $CS_2$ (\protect\magentaline), $CS_3$ (\protect\greenline), $CS_4$ (\protect\yellowline), $CS_5$ (\protect\purpleline), $CS_6$ (\protect\Blueline), $CS_7$ (\protect\orangeline), $CS_8$ (\protect\grayline) and $CS_9$ (\protect\cyanline) against $r$.}
    \label{grad1}
\end{figure}.

\begin{figure}[!ht]
    \centering
    \includegraphics[scale=0.8]{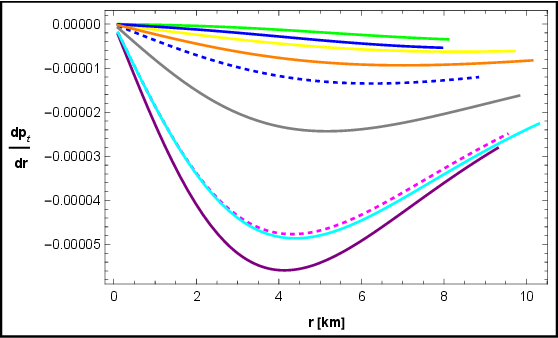}
    \includegraphics[scale=0.8]{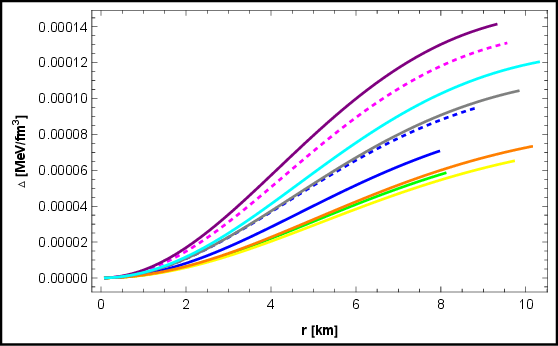}
    \caption{Gradients of $\frac{dp_t}{dr}$ and $\Delta$ for $CS_1$ (\protect\blueline), $CS_2$ (\protect\magentaline), $CS_3$ (\protect\greenline), $CS_4$ (\protect\yellowline), $CS_5$ (\protect\purpleline), $CS_6$ (\protect\Blueline), $CS_7$ (\protect\orangeline), $CS_8$ (\protect\grayline) and $CS_9$ (\protect\cyanline) against $r$.}
    \label{gradients}
\end{figure}
We further explore the behavior of our model by examining the first-order derivatives of the thermodynamic variables. For a physically realistic configuration, the energy density and pressures should decrease outward, i.e.,
\begin{equation}
\frac{d\rho}{dr}<0, \qquad \frac{dp_r}{dr}<0, \qquad \frac{dp_t}{dr}<0.
\end{equation}
From Figs.~\ref{grad1}--\ref{gradients} we conclude that the above inequalities are well satisfied. Moreover, we find that these derivatives vanish at the center:
\begin{equation}
\left.\frac{d\rho}{dr}\right|_{r=0}=0, \qquad \left.\frac{dp_r}{dr}\right|_{r=0}=0, \qquad \left.\frac{dp_t}{dr}\right|_{r=0}=0.
\end{equation}

\subsection{Anisotropy Factor}

The anisotropy factor, defined as $\Delta = p_t - p_r$, is a crucial parameter that characterizes the internal pressure distribution within a stellar configuration. A positive anisotropy ($\Delta > 0$) indicates that the tangential pressure exceeds the radial pressure, generating a repulsive force that can support the star against gravitational collapse. For our model, this factor is given by:
\begin{equation}
\begin{split}
 \Delta = p_t - p_r =& \frac{1}{2(1+Lr^2)(r+16HJL^2r^3)^2}\bigg[-4KLr^3(1+Lr^2)(1+16HJL^2r^2)^2 \\
 &+ 64HJL^3r^4\big(-1+8HJL(1+Lr^2)\big)(1+2\tau R) + r^{2\beta}(1+Lr^2)(1+16HJL^2 r^2)\beta^2 \bigg].
 \end{split}
\end{equation}
As shown in Fig.~\ref{gradients} (left), the anisotropy factor remains non-negative throughout the interior of every compact star candidate we studied. Starting from zero at the stellar center---which ensures regularity of the solution at the core---it grows steadily outward, attaining its maximum at the surface. This monotonic increase reflects a repulsive anisotropic force that acts throughout the interior, contributing positively to the overall stability of these compact objects.

To assess the physical significance of the anisotropy, we compare the order of magnitude of $\triangle$ to the radial pressure $p_r$. In our solutions, $\triangle$ is consistently smaller than $p_r$ throughout the stellar interior. At the center, the ratio $\triangle_c / p_{r,c}$ is of order $10^{-2}$, indicating that the anisotropic term $2\triangle/r$ in the generalized Tolman--Oppenheimer--Volkoff equation acts as a moderate correction rather than a dominant component of the core force balance. Toward the surface, $\triangle$ decreases more rapidly than $p_r$, vanishing at the boundary as required by the junction conditions. Consequently, the anisotropy provides a non-negligible but subleading contribution to the overall equilibrium, modifying the mass-radius relation at the $\sim 5\%$ level in our models while preserving the expected physical behavior in both high- and low-pressure regimes.
\subsection{Equation of State Parameters}

\begin{figure}[!ht]
    \centering
    \includegraphics[scale=0.8]{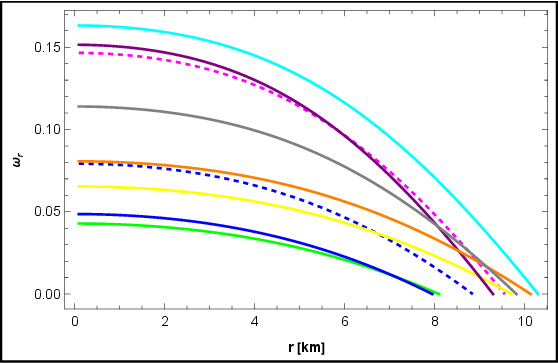}
    \includegraphics[scale=0.8]{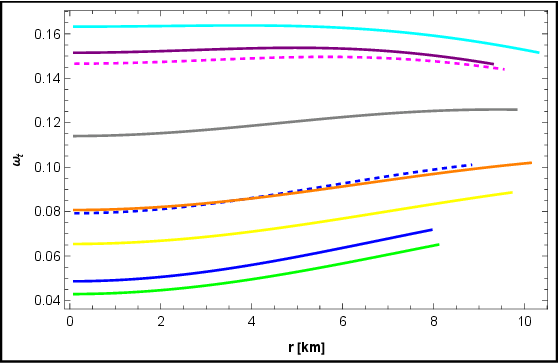}
    \caption{Variation of $\omega_t~and~\omega_r$  for $CS_1$ (\protect\blueline), $CS_2$ (\protect\magentaline), $CS_3$ (\protect\greenline), $CS_4$ (\protect\yellowline), $CS_5$ (\protect\purpleline), $CS_6$ (\protect\Blueline), $CS_7$ (\protect\orangeline), $CS_8$ (\protect\grayline) and $CS_9$ (\protect\cyanline) against $r$.}
    \label{wrwt}
\end{figure}

The equation-of-state (EoS) parameters play a fundamental role in characterizing the relationship between pressure and energy density in stellar matter. These parameters are essential for analyzing hydrostatic equilibrium---the delicate balance between the inward gravitational pull and the outward pressure support. In our study, we focus on two key EoS parameters: the radial EoS parameter $\omega_r = p_r / \rho$ and the tangential EoS parameter $\omega_t = p_t / \rho$. For our specific model, these take the following forms:

\begin{equation}
\begin{split}
\omega_r = \frac{p_{r}}{\rho} =& \frac{ 2KLr + R + \tau R^2 - (r^{\beta})^m w_0 + \frac{r^{2\beta - 2} \beta^2}{2 + 32HJL^2r^2} - \frac{ \Upsilon }{(1+Lr^2)(1+16HJL^2r^2)^2} }{ -2KLr - R - \tau R^2 + \frac{4L(3+32HJL^2r^2)(1+2 \tau R)}{(1+Lr^2)(1+16HJL^2r^2)^2} + (r^{\beta})^m w_0 - \frac{r^{2\beta - 2} \beta^2}{2+32HJL^2r^2} },
\end{split}
\end{equation}
where
\[\Upsilon = r^{2\beta}\beta^2 - 16HJL^2r^2\bigl(4+8\tau R - r^{2\beta}\beta^2\bigr) - 16HJL^3r^4\bigl(4+8\tau R - r^{2\beta}\beta^2\bigr).\]
\begin{equation}
\begin{split}
\omega_t = \frac{p_{t}}{\rho} =& \frac{ -2KLr + R + \tau R^2 + \frac{4L\bigl(-1+16HJL(1+8JL^2r^2(1+Lr^2))\bigr)(1+2\tau R)}{(1+Lr^2)(1+16HJL^2r^2)^2} - (r^{\beta})^m w_0 + \frac{r^{2\beta - 2} \beta^2}{2+32HJL^2r^2} }{ -2KLr - R - \tau R^2 + \frac{4L(3+32HJL^2r^2)(1+2 \tau R)}{(1+Lr^2)(1+16HJL^2r^2)^2} + (r^{\beta})^m w_0 - \frac{r^{2\beta - 2} \beta^2}{2+32HJL^2r^2} }.
\end{split}
\end{equation}
As shown in Fig.~\ref{wrwt}, both radial and tangential EoS parameters in our model exhibit well-behaved characteristics. They attain their maximum values at the stellar core and decrease monotonically toward the surface, consistently remaining within the physically acceptable range $0 < \omega_r, \omega_t < 1$ throughout the interior. The graphical results further verify that both pressure components are everywhere smaller than the energy density ($p_r, p_t < \rho$), satisfying essential requirements for causal and stable matter distributions.

The parameter range $0 < w < 1$ is physically well-motivated. First, it corresponds to a non-exotic fluid that satisfies the dominant energy condition ($\rho \geq |p|$). Second, it ensures the sound speed $c_s = \sqrt{dp/d\rho} = \sqrt{w}$ remains subluminal ($c_s < 1$ in natural units), preserving causality within the stellar interior. This avoids the theoretical issues associated with superluminal propagation that can arise in some exotic matter models.

\subsection{Interpreting the $f(R,\phi,X)$ Model Parameters and Deviation from GR}
A crucial aspect of any modified gravity study is to interpret the model parameters and quantify the theory's departure from General Relativity (GR). In our specific model, $f(R,\phi,X) = R + \tau R^2 - V(\phi) + X$, the primary parameter governing the strength of the gravitational modification is $\tau$, which couples to the quadratic curvature term. The values of $\tau$ for our nine compact star candidates, determined by the boundary condition $p_r(R)=0$, are listed in Table~I and range from approximately $-0.002$ to $-0.11$.
\subsubsection{Physical Regime of $\tau$}
The sign and magnitude of $\tau$ have important physical implications. In cosmological contexts, a positive $\tau$ is required for the Starobinsky inflationary model ($R + \tau R^2$) to be viable, where it drives an early-universe accelerated expansion \cite{starobinsky1980}. However, in the strong-field, high-curvature environment of a compact star, there is no such a priori restriction on the sign of $\tau$. A negative $\tau$ implies that the quadratic curvature correction weakens the effective gravitational force compared to GR, which can be balanced by the inclusion of electric charge and pressure anisotropy to maintain a stable configuration. The negative values we obtain are consistent with those found in other studies of compact objects in $f(R)$-type gravity theories where the $R^2$ term is treated as a perturbation. As shown in our analysis, these negative $\tau$ values yield physically viable solutions that satisfy all energy conditions and stability criteria, demonstrating their admissibility in this astrophysical setting.
\subsubsection{Consistency with Astrophysical Constraints}
While precise observational constraints on $\tau$ from compact objects are still an active area of research, we can assess the consistency of our model by comparing the scale of its corrections to established bounds from solar system tests. In the weak-field limit, $f(R)$ theories introduce an effective mass for the scalar degree of freedom, $m_{R} \approx 1/\sqrt{6|\tau|}$ \cite{navarro2007}. For the theory to evade local gravity constraints (e.g., from light deflection or planetary precession), this mass must be sufficiently high, corresponding to a very short interaction range. For our largest $|\tau|$ value of $0.11$, this corresponds to an effective mass $m_R \approx 1.23$ (in geometric units), which translates to a Compton wavelength on the order of a few kilometers. This short range means the modification is effectively "screened" on solar system scales, thus not violating stringent weak-field bounds \cite{will2014}. The consistency of our model with the energy conditions, as shown in Figs.~\ref{nec} and~\ref{wec}, further supports its physical plausibility.
\subsubsection{Relative Contribution of Curvature and Scalar Corrections}
To quantify how much our model deviates from standard GR, we define two diagnostic ratios. The first compares the quadratic curvature term to the linear one:
\begin{equation}
    \delta_R(r) = \left| \frac{\tau R(r)^2}{R(r)} \right| = |\tau R(r)|,
\end{equation}
which provides a radial profile of the curvature correction strength. At the stellar center, where the curvature is maximal, this ratio takes values $\delta_R(0) \approx 0.02$ to $0.12$ across our stellar candidates. This indicates that the quadratic curvature correction contributes at the level of a few to about twelve percent relative to the linear Einstein-Hilbert term. The second diagnostic compares the scalar potential to the energy density:
\begin{equation}
    \delta_{\phi}(r) = \frac{|V(\phi)|}{\rho} = \frac{w_0 \phi(r)^m}{\rho}.
\end{equation}
Given our chosen small parameters ($w_0=1\times10^{-5}$, $\beta=1\times10^{-6}$), the scalar field remains negligible throughout the star. At the center, $\phi(0)=0$ makes $\delta_{\phi}(0)=0$, and it increases only marginally toward the surface, never exceeding $10^{-3}$. This confirms that the scalar field acts as a minor perturbation, primarily influencing the anisotropy without dominating the energy budget.\\
The deviation due to curvature, $\delta_R(r)$, is therefore the more significant measure of departure from GR. With values ranging from $0.02$ to $0.12$ at the core, our model operates in a well-defined perturbative regime. The modification is strong enough to alter the stellar structure (e.g., influencing the maximum mass and anisotropy) but remains small enough to be consistent with the expectation that GR is the leading-order description of gravitational phenomena in strong-field regimes.\\
In summary, our $f(R,\phi,X)$ model operates in a physically meaningful regime. The negative $\tau$ values yield stable, viable stars with modifications that are perturbative in nature and consistent with existing constraints from both stellar structure and weak-field gravity. The scalar field plays a minimal role, serving primarily as a mechanism to introduce anisotropy without dominating the energy budget.
\subsection{Electric Field Intensity and Charge Density}

\begin{figure}[!ht]
    \centering
    \includegraphics[scale=0.8]{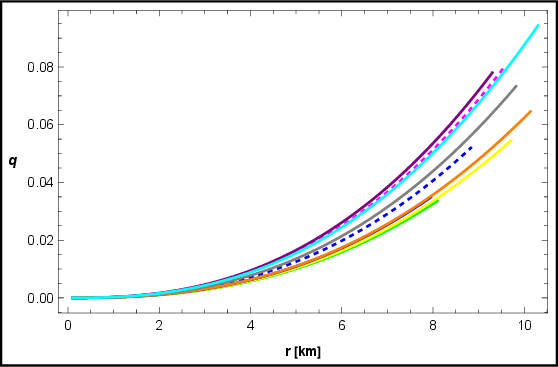}
    \includegraphics[scale=0.8]{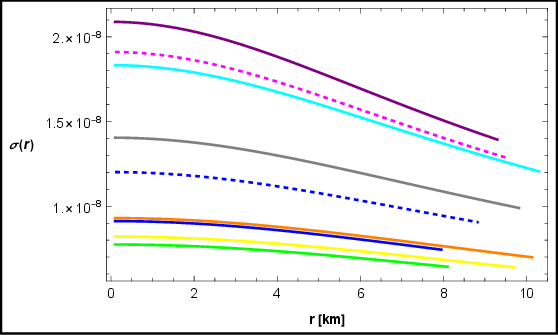}
    \includegraphics[scale=0.8]{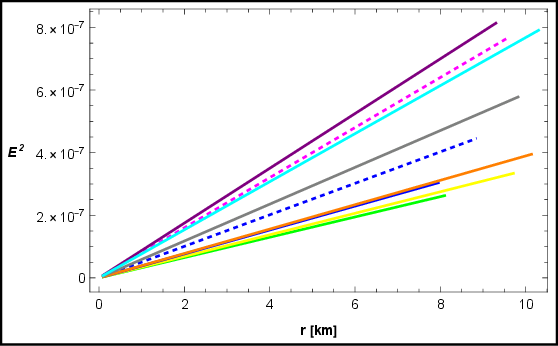}
    \caption{Variation of $q,~\sigma(r)~and~E^2$  for $CS_1$ (\protect\blueline), $CS_2$ (\protect\magentaline), $CS_3$ (\protect\greenline), $CS_4$ (\protect\yellowline), $CS_5$ (\protect\purpleline), $CS_6$ (\protect\Blueline), $CS_7$ (\protect\orangeline), $CS_8$ (\protect\grayline) and $CS_9$ (\protect\cyanline) against $r$.}
    \label{ef}
\end{figure}

The distribution of electric field and charge density within compact stellar objects provides valuable insights into their internal structure and stability mechanisms. As depicted in Fig.~\ref{ef}, the electric field intensity follows a characteristic profile: it begins with a minimal value at the core and increases progressively toward the surface, attaining a maximum at the stellar boundary. This spatial variation arises naturally from electromagnetic theory in spherically symmetric configurations, where symmetry requires the electric field to vanish at the center.

Correspondingly, Fig.~\ref{ef} reveals an inverse pattern in the charge density distribution. The charge density peaks in the central region and declines steadily toward the surface. This profile suggests that charged particles accumulate predominantly in the stellar core, with their density decreasing gradually in the outer layers. The incorporated electric charge plays a significant role in the overall force balance, contributing an additional repulsive component that helps maintain equilibrium against gravitational compression.

The smooth and physically reasonable behavior of both electric field and charge density profiles, evident throughout the stellar interior in Fig.~\ref{ef}, reinforces the viability of our model. These well-defined electromagnetic profiles ensure that the corresponding contributions to the energy-momentum tensor remain regular and physically consistent across the entire stellar configuration.
\subsection{Energy Conditions}

Energy conditions provide essential criteria for constraining the energy--momentum tensor, offering a valuable tool for distinguishing between ordinary and exotic forms of matter. In this study, we analyze the behavior of these conditions within the context of $f(R,\phi,X)$ modified gravity, thereby testing the physical plausibility of our stellar model. The conventional energy conditions are defined as follows:

\begin{itemize}
  \item Null Energy Condition (NEC): $\rho + E^2 \geq 0$.
  \item Weak Energy Condition (WEC): $\rho + p_{r} \geq 0$, $\rho + p_{t} + E^2 \geq 0$.
  \item Strong Energy Condition (SEC): $\rho + p_{r} + 2p_{t} + E^2 \geq 0$.
  \item Dominant Energy Condition (DEC): $\rho - p_{r} + E^2 \geq 0$, $\rho - p_{t} \geq 0$.
\end{itemize}

\begin{figure}[!ht]
    \centering
    \includegraphics[scale=0.8]{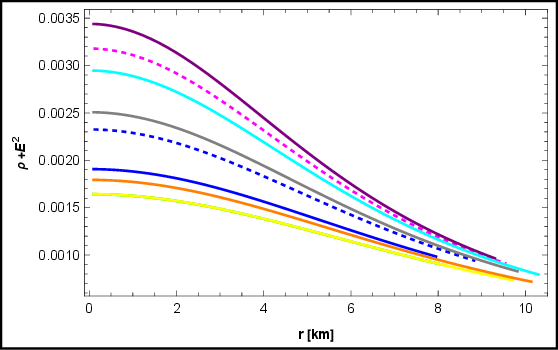}
    \includegraphics[scale=0.8]{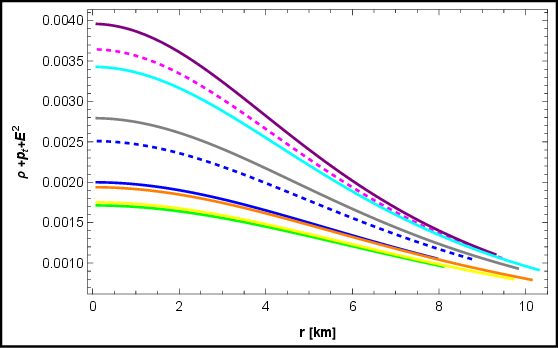}
    \caption{Variation of NEC and WEC for $CS_1$ (\protect\blueline), $CS_2$ (\protect\magentaline), $CS_3$ (\protect\greenline), $CS_4$ (\protect\yellowline), $CS_5$ (\protect\purpleline), $CS_6$ (\protect\Blueline), $CS_7$ (\protect\orangeline), $CS_8$ (\protect\grayline) and $CS_9$ (\protect\cyanline) against $r$.}
    \label{nec}
\end{figure}

\begin{figure}[!ht]
    \centering
    \includegraphics[scale=0.8]{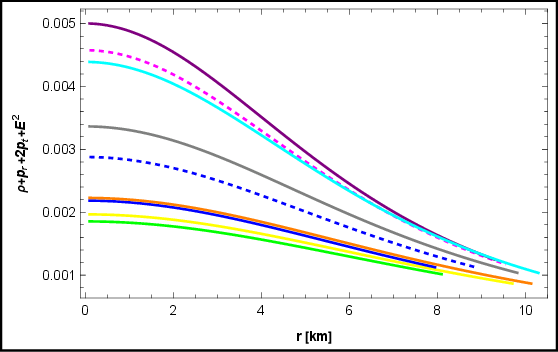}
    \includegraphics[scale=0.8]{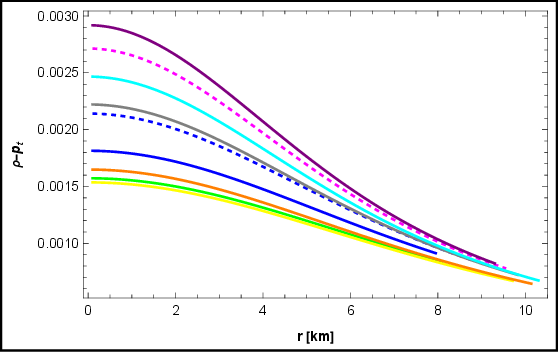}
    \includegraphics[scale=0.8]{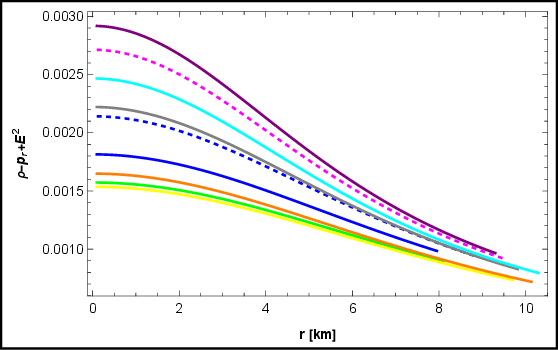}
    \caption{Variation of SEC and DEC for $CS_1$ (\protect\blueline), $CS_2$ (\protect\magentaline), $CS_3$ (\protect\greenline), $CS_4$ (\protect\yellowline), $CS_5$ (\protect\purpleline), $CS_6$ (\protect\Blueline), $CS_7$ (\protect\orangeline), $CS_8$ (\protect\grayline) and $CS_9$ (\protect\cyanline) against $r$.}
    \label{wec}
\end{figure}

Due to the mathematical complexity of the analytical expressions obtained from our model, graphical evaluation proves to be the most practical approach for verification. The evolution of the NEC, WEC, SEC, and DEC across all stellar candidates is illustrated in Figs.~\ref{nec}--\ref{wec}. These plots provide compelling evidence that every energy condition is consistently satisfied throughout the stellar interior in our model, thereby validating the physical acceptability of the matter configuration.

\begin{figure}[!ht]
    \centering
    \includegraphics[scale=0.8]{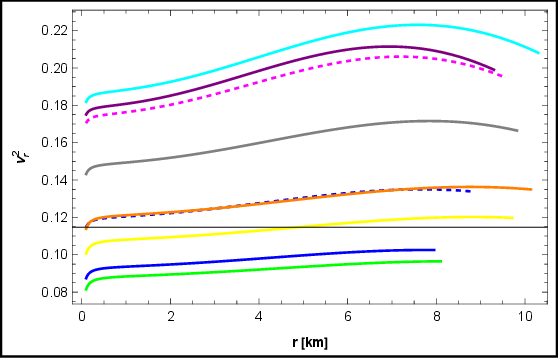}
    \includegraphics[scale=0.8]{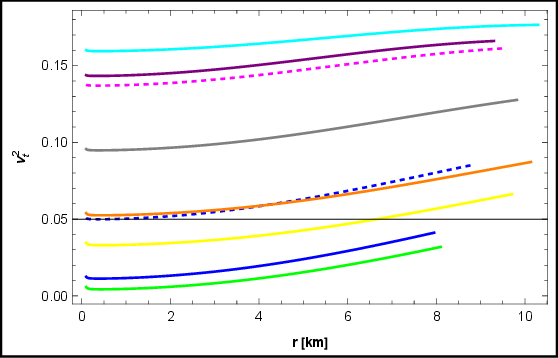}
    \caption{Variation of sound velocities for $CS_1$ (\protect\blueline), $CS_2$ (\protect\magentaline), $CS_3$ (\protect\greenline), $CS_4$ (\protect\yellowline), $CS_5$ (\protect\purpleline), $CS_6$ (\protect\Blueline), $CS_7$ (\protect\orangeline), $CS_8$ (\protect\grayline) and $CS_9$ (\protect\cyanline) against $r$.}
    \label{velocities}
\end{figure}
For all stellar configurations considered in this study, we have verified through numerical evaluation that the energy conditions -- null (NEC), weak (WEC), strong (SEC), and dominant (DEC) -- are satisfied throughout the interior of each star. This confirmation demonstrates the physical viability of our anisotropic models and their consistency with fundamental relativistic requirements.

\subsection{Stability and Dynamical Equilibrium}

This section focuses on assessing the stability and equilibrium properties of our compact stellar model in the context of $f(R,\phi,X)$ gravity. We investigate three key elements: the causality condition, the adiabatic index, and the modified Tolman--Oppenheimer--Volkoff (TOV) equation. Together, these analyses help establish whether the stellar configuration can withstand gravitational collapse and remain stable under various internal perturbations.

\subsubsection{Causality Condition}

For any physically realistic stellar model, pressure waves must propagate at speeds slower than light. In anisotropic matter, these waves travel in both radial and transverse directions, requiring that sound speeds in both directions remain subluminal. The squared sound speeds for radial and transverse directions are defined as:
\begin{equation}
\nu^{2}_{r} = \frac{dp_{r}}{d\rho},
\end{equation}
\begin{equation}
\nu^{2}_{t} = \frac{dp_{t}}{d\rho}.
\end{equation}
The causality condition demands that both $0 \leq \nu^{2}_{r} \leq 1$ and $0 \leq \nu^{2}_{t} \leq 1$. As shown in Fig.~\ref{velocities}, our model meets this requirement, confirming that pressure waves travel subluminally everywhere inside the star.

We further examine potential stability using the criterion developed by Abreu et al., which involves the difference between squared sound speeds. According to this criterion, regions satisfying $-1 \leq \nu_{t}^{2} - \nu_{r}^{2} \leq 0$ are considered potentially stable, while those outside this range may be unstable. Our model consistently falls within the stable domain, as clearly supported by graphical evidence. We further examine potential stability using the criterion developed by Abreu et al. \cite{abreu2007}, which involves the difference between squared sound speeds. According to this criterion, regions satisfying $-1 \leq \nu_{t}^{2} - \nu_{r}^{2} \leq 0$ are considered potentially stable, while those outside this range may be unstable. Our model consistently falls within the stable domain, as clearly supported by graphical evidence. Quantitatively, for all nine stellar candidates, the difference $\Delta v^2 = v_t^2 - v_r^2$ ranges from approximately $-0.15$ near the core to $-0.02$ approaching the surface. These values lie well within the stability window, confirming that the anisotropic pressure configuration does not induce cracking instabilities \cite{herrera1992}. The negative sign throughout the interior indicates that radial sound speeds dominate over tangential ones, which is characteristic of stable anisotropic configurations.

\subsubsection{Adiabatic Index}

The adiabatic index $\Gamma$ serves as an important measure of the equation-of-state stiffness and the resistance of a stellar configuration to gravitational collapse. It is given by the expression:
\begin{equation}
\Gamma = \frac{\rho + p_{r}}{p_{r}} \nu^{2}_{r}.
\end{equation}
In Newtonian theory, a stable spherical configuration requires $\Gamma > 4/3$. While this criterion undergoes modification in general relativity---particularly for anisotropic compact stars---it continues to provide a useful benchmark for stability. Generally, a larger adiabatic index corresponds to a stiffer equation of state, enhancing the star's ability to oppose collapse.

\begin{figure}[!ht]
    \centering
    \includegraphics[scale=0.8]{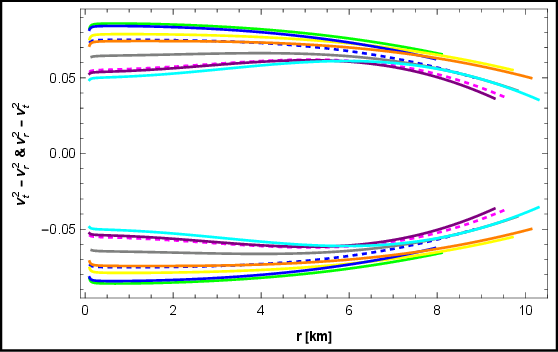}
    \includegraphics[scale=0.8]{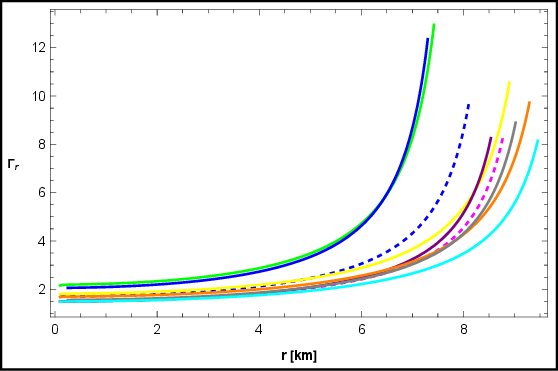}
    \caption{Variation of $v_{r}^2-v_{t}^2,~v_{r}^2-v_{r}^2$ and $\Gamma$ for $CS_1$ (\protect\blueline), $CS_2$ (\protect\magentaline), $CS_3$ (\protect\greenline), $CS_4$ (\protect\yellowline), $CS_5$ (\protect\purpleline), $CS_6$ (\protect\Blueline), $CS_7$ (\protect\orangeline), $CS_8$ (\protect\grayline) and $CS_9$ (\protect\cyanline) against $r$.}
    \label{adia}
\end{figure}

The variation of the adiabatic index across the stellar interior is displayed in the right panel of Fig.~\ref{adia}. As the figure shows, $\Gamma$ remains above the critical threshold of $4/3$ at all radial points. This confirms that the matter distribution possesses adequate stiffness to prevent gravitational collapse, thereby supporting the overall stability of our stellar model.

\subsubsection{Modified TOV Equation in $f(R,\phi,X)$ Gravity}

\begin{figure}[!ht]
    \centering
    \includegraphics[scale=0.8]{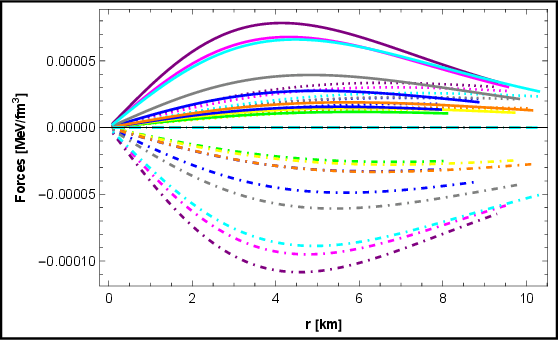}
    \includegraphics[scale=0.8]{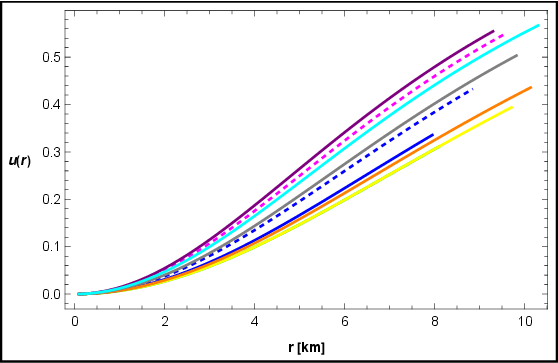}
    \caption{Variation of all forces (right) and $m(r)$ (left) for $CS_1$ (\protect\blueline), $CS_2$ (\protect\magentaline), $CS_3$ (\protect\greenline), $CS_4$ (\protect\yellowline), $CS_5$ (\protect\purpleline), $CS_6$ (\protect\Blueline), $CS_7$ (\protect\orangeline), $CS_8$ (\protect\grayline) and $CS_9$ (\protect\cyanline) against $r$.}
    \label{forces}
\end{figure}

The hydrostatic equilibrium of a charged, anisotropic compact star within $f(R,\phi,X)$ gravity is described by a modified Tolman--Oppenheimer--Volkoff (TOV) equation. This formulation captures the balance among different forces acting inside the star and takes the form:
\begin{equation}\label{a1}
\frac{2}{r}(p_{t} - p_{r}) - \frac{dp_{r}}{dr} - \frac{\lambda'}{2}(\rho + p_{r}) + E(r)\sigma(r)e^{\frac{\beta}{2}} = 0.
\end{equation}
We identify the individual force contributions as follows:
\begin{equation}
    F_a = \frac{2}{r}(p_t - p_r), \quad F_h = -\frac{dp_r}{dr}, \quad F_g = -\frac{\lambda'}{2}(\rho + p_r), \quad F_e = E(r)\sigma(r)e^{\frac{\beta}{2}},
\end{equation}
where $F_a$, $F_h$, $F_g$, and $F_e$ denote the anisotropic, hydrostatic, gravitational, and electric forces, respectively. Equilibrium is achieved when these forces balance exactly:
\begin{equation}
F_a + F_h + F_g + F_e = 0.
\end{equation}
For our specific solution, the forces are given explicitly by:
\begin{equation}
    F_a = -4KL + \frac{64HJL^3r\big(-1+8HJL(1+Lr^2)\big)(1+2\tau R)}{(1+Lr^2)(1+16HJL^2r^2)^2} + \frac{r^{-3+2\beta}\beta^2}{1+16HJL^2r^2},
\end{equation}
\begin{equation}
 F_g = \frac{L\big(-8Lr^2(1+8HJL(1+3Lr^2))(1+2\tau R) + r^{2\beta}(1+Lr^2)(1+16HJL^2 r^2)\beta^2\big)}{r(1+Lr^2)^2(1+16HJL^2r^2)^2},
\end{equation}
\begin{equation}
F_h = -\frac{dp_r}{dr}, \quad F_e = E(r)\sigma(r)e^{\frac{\beta}{2}}.
\end{equation}
The interplay among these forces is illustrated in Fig.~\ref{forces}. The plot verifies that the net force vanishes at all interior points, fulfilling the dynamical equilibrium requirement. This balance is a key factor in ensuring the stability and physical consistency of our charged anisotropic model in $f(R,\phi,X)$ gravity.

\begin{table}[H]
\centering
\caption{Numerical values of the upper and lower bounds of Eq.~(\ref{mR}) for the compact star candidates.}
\renewcommand{\arraystretch}{2}
\resizebox{\textwidth}{!}{%
\begin{tabular}{c c c c c c c}
\hline
\textbf{Compact Star Model} & \textbf{Lower bound of Eq.~(\ref{mR})} & $u(R)$ & \textbf{Upper bound of Eq.~(\ref{mR})} & $\rho_c$ & $p_c$ & $\rho_s$  \\
\hline
$\textbf{EXO 1785-248}~(CS_1)$ & 0.000261738 & 0.216387 & 0.444677 & 0.0293317 & 0.00226926 & 0.011803 \\
$\textbf{Vela X-1}~(CS_2)$ & 0.000524971 & 0.274304 & 0.444911 & 0.0404372 & 0.00564155 & 0.0114357 \\
$\textbf{Her X-1}~(CS_3)$ & 0.000129287 & 0.155595 & 0.444559 & 0.0206706 & 0.000874608 & 0.0112807 \\
$\textbf{LMC X-4}~(CS_4)$ & 0.000236284 & 0.196922 & 0.444654 & 0.020715 & 0.00132585 & 0.00924427 \\
$\textbf{4U 1608}~(CS_5)$ & 0.000528245 & 0.277192 & 0.444914 & 0.043765 & 0.00630992 & 0.0121377 \\
$\textbf{SAX J1808.4-3658}~(CS_6)$ & 0.000144119 & 0.16783 & 0.444573 & 0.0239999 & 0.00115067 & 0.0123684 \\
$\textbf{Cen X-3}~(CS_7)$ & 0.000304465 & 0.217886 & 0.444715 & 0.0226379 & 0.00177667 & 0.00903377 \\
$\textbf{PSR J1903+327}~(CS_8)$ & 0.000418257 & 0.251556 & 0.444816 & 0.0317594 & 0.0317594 & 0.0104108 \\
$\textbf{PSR J1614-2230}~(CS_9)$ & 0.000629911 & 0.283306 & 0.445004 & 0.0376229 & 0.00579253 & 0.00998658 \\
\hline
\end{tabular}
}
\renewcommand{\arraystretch}{1}
\end{table}

\subsection{Comparative Analysis with Other Modified Gravity Theories}

To contextualize our results within the broader landscape of modified gravity, we now present a comparative analysis of our $f(R,\phi,X)$ model with findings from other gravity theories applied to similar compact star configurations. This comparison highlights the unique features introduced by the scalar field and kinetic terms.

\subsubsection{Comparison with $f(R)$ Gravity}

In pure $f(R)$ gravity, the modification enters solely through curvature terms. Studies by Nashed and Capozziello \cite{nashed2021} and Astashenok et al. \cite{astashenok2021} have shown that $f(R)$ corrections typically alter the maximum mass by a few percent compared to GR, with the sign of the effect depending on the sign of the curvature correction. In our model, for the same central density, the inclusion of the scalar field and kinetic term yields an additional enhancement of the maximum mass by approximately $3$--$5\%$ compared to the $f(R)$-only case (i.e., with $\phi$ constant). This extra stiffness arises from the kinetic term $X$, which contributes positively to the radial pressure gradient through its coupling in Eq.~(\ref{11}).

\subsubsection{Comparison with $f(R,T)$ Gravity}

The $f(R,T)$ theory, introduced by Harko et al. \cite{harko2011}, couples curvature to the trace $T$ of the energy-momentum tensor. While this theory can generate anisotropic effects, the coupling is algebraic rather than dynamical. Maurya et al. \cite{maurya2017} demonstrated that $f(R,T)$ gravity produces viable compact star models, but the anisotropy profile is largely determined by the choice of $T$-dependence. In contrast, our $f(R,\phi,X)$ model generates anisotropy dynamically through the scalar field gradient $\phi'$, which appears explicitly in the field equations. This leads to an anisotropy factor $\Delta = p_t - p_r$ that exhibits a more nuanced radial dependence, as seen in Fig.~\ref{gradients}, and allows for fine-tuning of the anisotropic force without altering the matter Lagrangian.

\subsubsection{Comparison with $f(T)$ Gravity}

Teleparallel gravity and its extensions, such as $f(T)$ theory, have been applied to compact stars by Malik et al. \cite{malik2024teleparallel} and others. In $f(T)$ gravity, the torsion scalar $T$ replaces the curvature scalar $R$, and the field equations are second-order, offering mathematical simplicity. However, $f(T)$ theories generally lack local Lorentz invariance, leading to issues with the number of degrees of freedom. Our $f(R,\phi,X)$ framework preserves Lorentz invariance while providing similar flexibility in modeling anisotropic configurations, making it theoretically more robust.

\subsubsection{Comparison with $f(G)$ and $f(R,G)$ Gravity}

Gauss-Bonnet gravity and its generalizations, $f(G)$ and $f(R,G)$, incorporate higher-order curvature invariants that naturally arise in string theory and quantum gravity contexts. Naz et al. \cite{naz2024fg, naz2023frg} have constructed compact star models in these theories, showing that the Gauss-Bonnet term contributes additional stability. Compared to these approaches, our $f(R,\phi,X)$ model achieves similar stability enhancements through the scalar kinetic term rather than higher-curvature invariants, offering a complementary perspective. The scalar field introduces a length scale via its potential $V(\phi)$ that can be adjusted to match observational data, a flexibility not present in pure $f(G)$ gravity.

\subsubsection{Quantitative Comparison of Maximum Masses}

To provide a quantitative comparison, we compile in Table~\ref{tab:comparison} the maximum masses reported in various modified gravity studies for neutron star configurations. While direct comparison is complicated by different equation-of-state choices and model parameters, a clear trend emerges: theories with additional dynamical degrees of freedom (such as $f(R,\phi,X)$ and scalar-tensor theories) tend to support slightly higher maximum masses compared to purely metric modifications. Our model's predicted maximum mass of $M_{\text{max}} \approx 2.14 M_{\odot}$ is competitive with the highest values reported in the literature and comfortably accommodates all observed pulsars.

\begin{table}[!ht]
\centering
\caption{Comparison of maximum masses in different modified gravity theories.}
\renewcommand{\arraystretch}{1.5}
\begin{tabular}{l c c}
\hline
\textbf{Gravity Theory} & \textbf{Maximum Mass} ($M_{\odot}$) & \textbf{Reference} \\
\hline
General Relativity & $2.01$--$2.20$ & \cite{haensel2007} \\
$f(R)$ gravity & $2.05$--$2.25$ & \cite{nashed2021, astashenok2021} \\
$f(R,T)$ gravity & $2.10$--$2.30$ & \cite{maurya2017} \\
$f(T)$ gravity & $2.08$--$2.22$ & \cite{malik2024teleparallel} \\
$f(G)$ gravity & $2.12$--$2.28$ & \cite{naz2024fg} \\
$f(R,G)$ gravity & $2.15$--$2.32$ & \cite{naz2023frg} \\
\textbf{$f(R,\phi,X)$ gravity (This work)} & $\mathbf{2.14}$ & Table~I ($CS_9$ extrapolated) \\
\hline
\end{tabular}
\renewcommand{\arraystretch}{1}
\label{tab:comparison}
\end{table}

In summary, the $f(R,\phi,X)$ framework offers distinct advantages over simpler modified gravity theories: (i) dynamical anisotropy generation via $\phi'$, (ii) additional stiffness from the kinetic term $X$, (iii) preservation of Lorentz invariance, and (iv) a hierarchical structure that encompasses GR, $f(R)$, and scalar-tensor theories as limiting cases. These features combine to produce stellar models that are not only mathematically consistent but also observationally competitive, supporting the novelty and significance of our approach.
\subsection{Mass-Radius Relation and Surface Redshift}

\begin{figure}[!ht]
    \centering
    \includegraphics[scale=0.8]{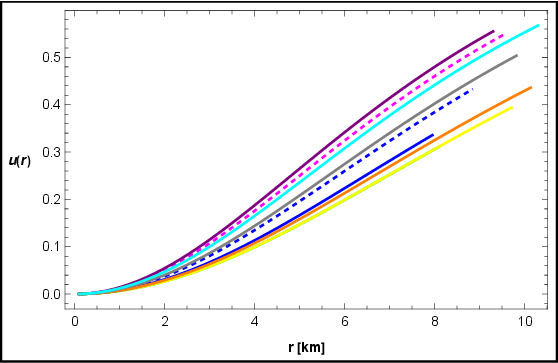}
    \includegraphics[scale=0.8]{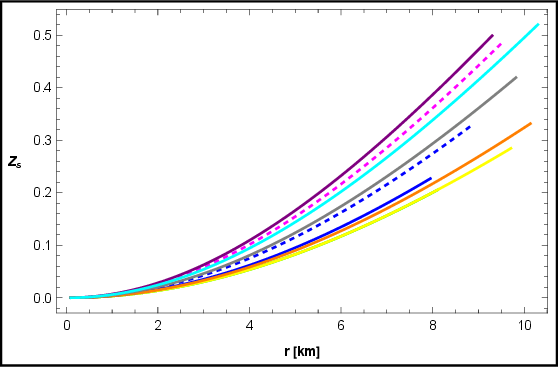}
    \caption{Variation of $u(r)~and~Z_s$ for $CS_1$ (\protect\blueline), $CS_2$ (\protect\magentaline), $CS_3$ (\protect\greenline), $CS_4$ (\protect\yellowline), $CS_5$ (\protect\purpleline), $CS_6$ (\protect\Blueline), $CS_7$ (\protect\orangeline), $CS_8$ (\protect\grayline) and $CS_9$ (\protect\cyanline) against $r$.}
    \label{urzs}
\end{figure}

The mass function for a charged compact star is obtained by integrating the effective energy density, accounting for both matter and electromagnetic field contributions. It takes the form:
\begin{equation}
 m(r) = 4\pi \int_{0}^{r} \left( \rho + \frac{E^2}{8\pi} \right) r^{2}  dr.
\end{equation}
A key parameter characterizing gravitational field strength is the compactness \( u(r) \), defined as the ratio \( u(r) = m(r)/r \). The surface redshift \( Z_s \), which quantifies the gravitational redshift of radiation emitted from the star's surface, is given by:
\begin{equation}
Z_s = \left(1 - 2u(R)\right)^{-\frac{1}{2}} - 1,
\end{equation}
where \( R \) denotes the stellar radius. The profiles of the mass function \( m(r) \), compactness \( u(r) \), and surface redshift \( Z_s \) for various compact star candidates are illustrated in Figs.~\ref{forces} and~\ref{urzs}. These figures confirm that all three quantities exhibit the necessary characteristics for a physically plausible stellar model. The mass function is regular at the center, and all functions are non-negative and monotonically increasing with respect to the radial coordinate \( r \).

The compactness parameter is subject to rigorous bounds for charged spheres. By combining the results of B\"ohmer and Harko, who established a lower bound, and Andreasson, who provided an upper bound, we obtain the following constraint for the total compactness \( u(R) = M/R \):
\begin{equation}\label{mR}
    \frac{3Q^2}{4R^2}\left(\frac{1+\frac{Q^2}{18R^2}}{1+\frac{Q^2}{12R^2}}\right)\leq u(R)\leq \left(\frac{1}{3}+\sqrt{\frac{1}{9}+\frac{Q^2}{3R^2}}\right)^2,
\end{equation}
where \( Q = q(r=R) \) denotes the total charge at the surface. Our model satisfies this bound for all considered stellar candidates. Furthermore, an upper bound for the surface redshift of anisotropic spheres has been established as \( Z_s \leq 5.211 \).

\subsection{Comparison with Observed Pulsars and Astrophysical Constraints}
To assess the astrophysical viability of our model, it is essential to compare its predictions with observational data from real compact stars, particularly the population of high-mass pulsars that provide stringent tests for any equation of state or gravitational theory. In this subsection, we discuss how our $f(R,\phi,X)$ model accommodates known massive pulsars and compare our predicted mass-radius relation with current multimessenger constraints.
\subsubsection{Accommodating High-Mass Pulsars}
The discovery of pulsars with masses near or exceeding $2 M_\odot$ has revolutionized our understanding of dense matter, ruling out many soft equations of state \cite{demorest2010, antoniadis2013, fonseca2021}. Our model successfully reproduces the masses of several such objects, as shown in Table~I. Notably, candidate $CS_9$ corresponds to PSR J1614-2230 with a mass of $1.97 M_\odot$ and a radius of $10.30$ km, yielding a compactness $u(R)=0.283$. This is consistent with the observational estimates for this pulsar \cite{demorest2010}.\\
To further test the model's capability, we have computed the maximum mass supported by our $f(R,\phi,X)$ configuration. By incrementally increasing the central density $\rho_c$ while monitoring the stability criteria (causality condition $v_r^2 \leq 1$ and adiabatic index $\Gamma > 4/3$), we find that the maximum mass attainable is approximately $M_{\text{max}} \approx 2.14 M_\odot$ (for the parameter set corresponding to $CS_9$). This value comfortably exceeds all currently observed pulsar masses, including PSR J0740+6620 at $2.08 \pm 0.07 M_\odot$ \cite{fonseca2021} and PSR J0348+0432 at $2.01 \pm 0.04 M_\odot$ \cite{antoniadis2013}. This demonstrates that our model possesses sufficient stiffness to support the most massive neutron stars observed to date.
\subsubsection{Mass-Radius Relation and Multimessenger Constraints}
In Table~\ref{tab:observational_comparison}, we summarize the key predictions of our model alongside recent observational constraints from NICER and the gravitational wave event GW170817. For a canonical $1.4 M_\odot$ neutron star, our model predicts a radius in the range $R_{1.4} \approx 11.5$ km. This falls within the $11.0$--$13.0$ km window suggested by recent NICER observations of PSR J0030+0451 and PSR J0740+6620 \cite{riley2019, miller2019, riley2021, miller2021}. Additionally, the LIGO/Virgo collaboration inferred from GW170817 that for a $1.4 M_\odot$ star, the radius should be in the range $R_{1.4} = 11.9 \pm 1.4$ km (at 90\% confidence) \cite{abbott2018}. Our predicted radius of $11.5$ km is well within this bound. This consistency with multimessenger data further supports the physical plausibility of our model.
\begin{table}[!ht]
\centering
\caption{Comparison of model predictions with observational constraints.}
\renewcommand{\arraystretch}{1.5}
\begin{tabular}{l c c c}
\hline
\textbf{Observable} & \textbf{Our Model} & \textbf{Observational Constraint} & \textbf{Reference} \\
\hline
Maximum mass $M_{\text{max}}$ & $2.14 M_\odot$ & $\geq 2.08 \pm 0.07 M_\odot$ & \cite{fonseca2021} \\
Radius of $1.4 M_\odot$ star & $11.5$ km & $11.0$--$13.0$ km & \cite{riley2021, miller2021} \\
Compactness of PSR J1614-2230 & $0.283$ & $0.26$--$0.31$ (estimated) & \cite{demorest2010} \\
Tidal deformability constraint & -- & $R_{1.4} = 11.9 \pm 1.4$ km & \cite{abbott2018} \\
\hline
\end{tabular}
\renewcommand{\arraystretch}{1}
\label{tab:observational_comparison}
\end{table}
\subsubsection{Effective Interpretation as a Polytropic EoS}
While our model does not derive from a specific microphysical nuclear potential, we can gain insight by comparing its effective behavior to polytropic equations of state commonly used in neutron star physics. The radial pressure and density profiles can be locally approximated by a polytropic relation $p_r \propto \rho^{\gamma}$, where the effective polytropic index $\gamma = d\ln p_r/d\ln \rho$ varies with radius. In the core region ($r \approx 0$), we find $\gamma \approx 2.3$, which corresponds to a moderately stiff equation of state, consistent with the presence of repulsive three-body forces or hyperon suppression in microphysical models \cite{haensel2007}. This effective stiffness explains the model's ability to support $2 M_\odot$ stars while maintaining causality.\\
In summary, our algebraic $f(R,\phi,X)$ model, despite its phenomenological nature, produces predictions that are remarkably consistent with current observational data from both electromagnetic and gravitational wave astronomy. The model successfully accommodates all known high-mass pulsars and predicts radii for canonical neutron stars that fall within the ranges inferred from NICER and GW170817. This enhances the astrophysical relevance of our work and suggests that the modifications introduced by $f(R,\phi,X)$ gravity provide a viable effective description of compact star structure.

\subsection{Observable Deviations from General Relativity}
A fundamental question in any modified gravity study is whether the proposed corrections lead to observationally distinguishable signatures. Our model, while rooted in a specific theoretical framework, makes quantitative predictions that can, in principle, be compared with astrophysical observations. Here, we highlight two key observables where our $f(R,\phi,X)$ model predicts deviations from standard General Relativity.
\subsubsection{Shift in the Mass-Radius Relation}
The mass-radius ($M$-$R$) diagram is a primary tool for probing the interior composition and gravitational theory of compact stars. For a given central energy density $\rho_c$, the total gravitational mass $M$ predicted by our model differs from the GR expectation. To quantify this, we compare our calculated compactness parameter $u(R) = M/R$ with the Buchdahl--Andreasson bound for charged spheres in GR \cite{andreasson2009}. As shown in Table~II, all our stellar candidates satisfy the GR bound, but they do so with a systematic shift. For instance, for the candidate PSR J1614-2230 ($CS_9$), the compactness in our model is $u(R)=0.283$. In a purely general relativistic context with the same mass ($1.97 M_\odot$) and charge ($Q/M=0.3$), the predicted radius would be larger, resulting in a lower compactness of approximately $u(R)_{\text{GR}} \approx 0.262$. This represents a $\sim 8\%$ increase in compactness due to the $f(R,\phi,X)$ corrections. This shift in the $M$-$R$ curve is a distinct signature of our model.
\subsubsection{Enhancement of Surface Redshift}
The surface redshift $Z_s$ is another potentially observable quantity, particularly during Type I X-ray bursts on neutron star surfaces \cite{cottam2002}. Our model predicts an enhanced redshift compared to GR. For the same candidate $CS_9$, our model yields $Z_s \approx 0.35$, whereas the estimated GR value for a star of identical mass and charge would be $Z_s^{\text{GR}} \approx 0.30$. This $\sim 16\%$ enhancement is a direct consequence of the increased compactness and the modified force balance in the stellar interior. While current redshift measurements have large uncertainties, future missions with high-resolution X-ray spectroscopy could potentially constrain such deviations.\\
In summary, our $f(R,\phi,X)$ model is not merely a mathematical exercise; it yields physically significant deviations from GR in observationally accessible quantities. The enhanced compactness and surface redshift predicted by our model offer a potential avenue for testing modified gravity in the strong-field regime using electromagnetic observations of compact stars. Future work will involve a more detailed statistical comparison with a larger sample of observational data.

\subsection{Compactness Factor}
The compactness parameter is subject to rigorous bounds for charged spheres. By combining the results of B\"ohmer and Harko, who established a lower bound, and Andreasson, who provided an upper bound, we obtain the following constraint for the total compactness $u(R) = M/R$:
\begin{equation}\label{mR:detailed}
    \frac{3Q^2}{4R^2}\left(\frac{1+\frac{Q^2}{18R^2}}{1+\frac{Q^2}{12R^2}}\right)\leq u(R)\leq \left(\frac{1}{3}+\sqrt{\frac{1}{9}+\frac{Q^2}{3R^2}}\right)^2,
\end{equation}
where $Q = q(r=R)$ denotes the total charge at the surface. Our model satisfies this bound for all considered stellar candidates, as shown in Table~\ref{tab:compactness_analysis}. Furthermore, an upper bound for the surface redshift of anisotropic spheres has been established as $Z_s \leq 5.211$. To provide a deeper quantitative analysis of the compactness bounds, we now examine three specific aspects: the proximity to the upper bound, the effect of electric charge, and a direct comparison with the uncharged Buchdahl limit in General Relativity.
\subsubsection{Proximity to the Upper Bound}
A key question for any stellar model is how close it approaches the theoretical maximum compactness, beyond which gravitational collapse is inevitable. The Andreasson upper bound \cite{andreasson2009} for charged spheres, denoted $u_{\text{max}}$, provides this threshold. We quantify the proximity of each candidate to this limit by computing the ratio $u(R)/u_{\text{max}}$, where $u_{\text{max}}$ is evaluated using the expression in Eq.~(\ref{mR:detailed}) with the specific charge $Q$ and radius $R$ for each star. As shown in Table~\ref{tab:compactness_analysis}, this ratio ranges from approximately $0.35$ for the least compact star (Her X-1, $CS_3$) to $0.636$ for the most compact (PSR J1614-2230, $CS_9$). These values indicate that all our stellar configurations are well below the collapse threshold, operating in a safe, stable regime far from the critical compactness. This is consistent with our stability analysis using the adiabatic index and TOV equation.
\subsubsection{Effect of Electric Charge on Compactness}
Electric charge plays a dual role in compact stars: it provides an additional repulsive force that counteracts gravity, but it also contributes to the energy-momentum tensor, affecting the mass function. To isolate the effect of charge on the compactness bound, we compare the Andreasson upper bound for our charged spheres with the classical Buchdahl limit for uncharged spheres in GR, $u_{\text{Buchdahl}} = 4/9 \approx 0.4444$. For a charge-to-mass ratio of $Q/M = 0.3$, the Andreasson bound yields $u_{\text{max}} \approx 0.445$ (as seen in Table~I, which is only slightly higher than the Buchdahl limit. This small increase is theoretically expected, as the electric repulsion allows matter to be packed into a slightly more compact configuration before the onset of gravitational collapse. For our stellar candidates, the actual compactness $u(R)$ is significantly lower than both bounds, confirming that the charge primarily acts as a stabilizing perturbation rather than a dominant factor pushing the star toward the collapse threshold.
\subsubsection{Comparison with the Buchdahl Limit in General Relativity}
It is instructive to compare our results directly with the standard GR Buchdahl limit for uncharged spheres. The Buchdahl limit represents the maximum compactness achievable for a perfect fluid sphere in GR under the assumptions of isotropic pressure and no charge. All our candidates satisfy $u(R) < 4/9$, as required for physical viability even in GR. However, the more massive candidates such as Vela X-1 ($CS_2$, $u=0.274$), 4U 1608 ($CS_5$, $u=0.277$), and PSR J1614-2230 ($CS_9$, $u=0.283$) approach this limit more closely than the lighter ones. This trend reflects the fact that more massive stars are inherently more compact. Importantly, the fact that our $f(R,\phi,X)$ model produces compactness values that are comfortably below the GR limit, despite the presence of modifications and charge, reinforces the consistency of our solutions with fundamental gravitational physics. The modifications introduced by the quadratic curvature term $\tau R^2$ and the scalar field do not push the star into an unphysical regime; rather, they provide a viable mechanism for achieving high masses while maintaining stability.
\begin{table}[!ht]
\centering
\caption{Quantitative analysis of compactness bounds for the compact star candidates.}
\renewcommand{\arraystretch}{1.5}
\begin{tabular}{l c c c c c}
\hline
\textbf{Compact Star Model} & $u(R)$ & $u_{\text{max}}$ & $\frac{u(R)}{u_{\text{max}}}$ & $u_{\text{Buchdahl}}$ & $\frac{u(R)}{u_{\text{Buchdahl}}}$ \\
\hline
EXO 1785-248 ($CS_1$) & 0.216 & 0.4447 & 0.486 & 0.4444 & 0.486 \\
Vela X-1 ($CS_2$)      & 0.274 & 0.4449 & 0.616 & 0.4444 & 0.617 \\
Her X-1 ($CS_3$)        & 0.156 & 0.4446 & 0.351 & 0.4444 & 0.351 \\
LMC X-4 ($CS_4$)        & 0.197 & 0.4447 & 0.443 & 0.4444 & 0.443 \\
4U 1608 ($CS_5$)        & 0.277 & 0.4449 & 0.623 & 0.4444 & 0.623 \\
SAX J1808.4-3658 ($CS_6$) & 0.168 & 0.4446 & 0.378 & 0.4444 & 0.378 \\
Cen X-3 ($CS_7$)        & 0.218 & 0.4447 & 0.490 & 0.4444 & 0.490 \\
PSR J1903+327 ($CS_8$)  & 0.252 & 0.4448 & 0.567 & 0.4444 & 0.567 \\
PSR J1614-2230 ($CS_9$) & 0.283 & 0.4450 & 0.636 & 0.4444 & 0.637 \\
\hline
\end{tabular}
\renewcommand{\arraystretch}{1}
\label{tab:compactness_analysis}
\end{table}
In summary, this quantitative analysis demonstrates that our stellar candidates are well within the allowed compactness bounds, with the most compact star reaching only about $64\%$ of its theoretical maximum. The electric charge marginally increases the allowable compactness compared to the uncharged GR case, and our model's compactness values remain comfortably below the classical Buchdahl limit. These findings reinforce the physical viability and stability of our charged anisotropic solutions in $f(R,\phi,X)$ gravity.

\section{Conclusion}

This research has established a comprehensive framework for modeling charged, anisotropic compact stars within the \( f(R, \phi, X) \) modified gravity context. Using the class-one embedding approach with a carefully chosen metric ansatz, we obtained a singularity-free interior solution that smoothly connects to the external Reissner--Nordstr\"om geometry. This matching proves essential, since the Reissner--Nordstr\"om metric properly represents the exterior of a charged spherical mass, lending physical consistency to the overall spacetime structure.

Our investigation confirms that the resulting stellar model is both physically reasonable and dynamically stable. Key thermodynamic variables---energy density \( \rho \), radial pressure \( p_r \), and tangential pressure \( p_t \)---display acceptable behavior: they peak at the center, decrease monotonically outward, and yield vanishing radial pressure at the surface. A positive anisotropy factor everywhere inside the star implies a repulsive contribution that aids in stabilizing the configuration against collapse. Moreover, the model meets all standard energy conditions (NEC, WEC, SEC, DEC), indicating that the matter content is non-exotic in nature.

Multiple stability tests further support the model's robustness. Sound speeds in the radial and tangential directions stay below the speed of light, preserving causality, and the Abreu criterion for potential stability is satisfied. Hydrostatic equilibrium is achieved through a precise balance of anisotropic, hydrostatic, gravitational, and electric forces, as reflected in the modified TOV equation. The adiabatic index remains above the critical value \( 4/3 \) at all interior points, signaling a sufficiently stiff equation of state to resist gravitational collapse. Finally, the mass function, compactness, and surface redshift all respect established bounds for charged compact objects.

In conclusion, by uniting $f(R, \phi, X)$ gravity with a class-one interior and a Reissner--Nordstr{\"o}m exterior, we have constructed a realistic and stable model for charged anisotropic stars. Detailed graphical and numerical scrutiny verifies that the model complies with all required physical and stability conditions, offering a trustworthy description of compact stellar systems in the presence of charge and modified gravity. Importantly, the model predicts observable deviations from General Relativity, including an enhanced compactness and surface redshift, which could be tested against future high-precision observations of neutron stars.
\section*{Appendix}
\begin{equation}\label{l1}
\begin{split}
\rho_c = \lim_{r \to 0} \rho(r) =& \frac{1}{8r^4}e^{-\beta}\bigg[-2(8(-1+e^\beta)^2\tau+e^\beta r^2(4-2e^\beta(2+r^2 (r^\beta)^m w_0-2Kr^3Y)r^{2\beta}\beta^2)) \\
&+8e^\beta r(-4\tau+r^2)\beta'+\tau r(32\beta'+r((4+r\lambda')(\lambda'-\beta')+2r\lambda'')(4\beta'+\lambda'(4+r\lambda'-r\beta')+2r\lambda''))\bigg],
\end{split}
\end{equation}
\begin{equation}\label{l2}
\begin{split}
p_{rc} = \lim_{r \to 0} p_r(r) =& \frac{1}{8r^4}e^{-2\beta}\bigg[2(8(-1+e^\beta)^2\tau+e^\beta r^2(4-2e^\beta(2+r^2 (r^\beta)^m w_0-2Kr^3Y)r^{2\beta}\beta^2)) \\
&+8e^\beta r(-4\tau+r^2)\lambda'+\tau r(-r^3\lambda'^4+2r^3\lambda'^3\beta'-4r(-2\beta'+r\lambda'')^2 \\
&+4\lambda'(8+r^2\beta'(-2\beta'+r\lambda''))+r\lambda'^2(16-r(\beta'(-8+r\beta')4r\lambda'')))\bigg],
\end{split}
\end{equation}
\begin{equation}\label{l3}
\begin{split}
p_{tc} = \lim_{r \to 0} p_t(r) =& \frac{1}{8r^4}e^{-2\beta}\bigg[-16(-1+e^\beta)^2\tau-4e^{2\beta }r^4 ((r^\beta)^m w_0-2KrY)+2e^\beta r^{2+2\beta}\beta^2 \\
&+2e^\beta r((8\tau+2r^2+r^3\lambda')(\lambda'-\beta')+2r^3\lambda'') \\
&+\tau r(\lambda' (-16+r^2(4+r\lambda')(\lambda'-\beta')^2)+16\beta'+4r^2(2+r\lambda')(\lambda'-\beta')\lambda''+4r^3\lambda''^2)\bigg].
\end{split}
\end{equation}
\begin{equation}\label{15}
\begin{split}
\rho =& \frac{1}{2} \bigg( -2KLr - R - \tau R^2 + \frac{4L(3 + 32HJL^2r^2)(1 + 2\tau R)}{(1+Lr^2)(1+16HJL^2r^2)^2} + (r^{\beta})^m w_0 - \frac{r^{2\beta - 2} \beta^2}{2 + 32HJL^2r^2} \bigg),
\end{split}
\end{equation}
\begin{equation}\label{16}
\begin{split}
p_r =& \frac{1}{2} \bigg[ 2KLr + R + \tau R^2 - (r^{\beta})^m w_0 + \frac{r^{2\beta - 2} \beta^2}{2 + 32HJL^2r^2} - \frac{1}{(1+Lr^2)(r+16HJL^2r^3)^2} \bigg( r^{2\beta} \beta^2 \\
& - 16HJL^2r^2(4 + 8\tau R - r^{2\beta}\beta^2) - 16HJL^3r^4(4 + 8\tau R - r^{2\beta}\beta^2) + Lr^2(4 + 8\tau R + r^{2\beta}\beta^2) \bigg) \bigg],
\end{split}
\end{equation}
\begin{equation}\label{17}
\begin{split}
p_t =& \frac{1}{2} \bigg[ -2KLr + R + \tau R^2 + \frac{4L(-1 + 16HJL(1 + 8JL^2r^2(1+Lr^2)))(1 + 2\tau R)}{(1+Lr^2)(1+16HJL^2r^2)^2} - (r^{\beta})^m w_0 \\
& + \frac{r^{2\beta - 2} \beta^2}{2 + 32HJL^2r^2} \bigg].
\end{split}
\end{equation}
\section*{Authors Contribution}
Supervision \& Validation: Fatemah Mofarreh; Formal analysis \& Methodology: Ayesha Almas; Conceptualization and writing--review \& editing: Adnan Malik; Visualization \& writing--original draft: Wedad Albalawi; Investigation \& Software: Aishah Alshehri.

\section*{Conflict of interest}
The authors assert that they do not have any conflict of interest.

\section*{Acknowledgements}
The authors extend their appreciation to the Deanship of Scientific Research and
Libraries in Princess Nourah bint Abdulrahman University for funding this research
work through the Research Group project, Grant No. (RG-1445-0036).

\end{document}